\RequirePackage{fix-cm}
\documentclass{svjour3}                     
\smartqed  
\RequirePackage[T1]{fontenc}

\RequirePackage{graphicx}
\RequirePackage{mathptmx}      
\RequirePackage{flushend}
\RequirePackage[numbers,sort&compress]{natbib}
\RequirePackage[colorlinks,citecolor=blue,urlcolor=blue,linkcolor=blue]{hyperref}
\usepackage{pdflscape}
\usepackage{amsmath, amssymb, graphics}
\newcommand{\mathsym}[1]{{}}
\newcommand{\unicode}[1]{{}}
\usepackage{textcomp}
\usepackage{eurosym}
\usepackage{widetext}
\usepackage{amsfonts}
\usepackage{array}
\usepackage{bm}
\usepackage{helvet}

\usepackage{graphicx}
\usepackage{subfigure}
\usepackage{bm, array}
\usepackage[T1]{fontenc}
\def\be{\begin{equation}}
\def\ee{\end{equation}}
\def\beq{\begin{eqnarray}}
\def\eeq{\end{eqnarray}}

\newcommand{\ben}{\begin{eqnarray}}
\newcommand{\een}{\end{eqnarray}}

\usepackage{color}
\usepackage{enumerate}
\usepackage{soul}
\usepackage[normalem]{ulem}
\usepackage{color}
\usepackage{color}
\usepackage{mathtools, cuted}

\begin{document}

\title{Stability of a modified Jordan-Brans-Dicke theory in the dilatonic frame}

\titlerunning{Stability of a modified Jordan-Brans-Dicke theory in the dilatonic frame} 

\author{Genly Leon \and Andronikos Paliathanasis \and Luisberis Velazquez Abad 
}

\institute{G. Leon \at
               Departamento  de  Matem\'aticas,  Universidad  Cat\'olica  del  Norte, Avda. Angamos  0610,  Casilla  1280  Antofagasta,  Chile.\\
\email{\href{mailto:genly.leon@ucn.cl}{genly.leon@ucn.cl} }
           \and
           Andronikos Paliathanasis \at
              Instituto de Ciencias F\'{\i}sicas y Matem\'aticas, Universidad Austral de Chile, Valdivia, Chile.\\
							Institute of Systems Science, Durban University of Technology, PO Box 1334, Durban 4000, South
Africa. \\
\email{\href{anpaliat@phys.uoa.gr}{anpaliat@phys.uoa.gr}}
           \and
           Luisberis Velazquez Abad \at
              Departamento de F\'isica, Universidad Cat\'olica del Norte, Avda. Angamos 0610, Casilla 1280 Antofagasta, Chile. \\
\email{\href{lvelazquez@ucn.cl}{lvelazquez@ucn.cl} }}

\date{Received: date / Accepted: date}

\maketitle
\begin{abstract}
We investigate the Jordan-Brans-Dicke action in the cosmological scenario of FLRW spacetime with zero spatially curvature and with an extra scalar field minimally coupled to gravity as matter source. The field equations are studied in two ways. The method of group invariant transformations, i.e., symmetries of differential equations apply in order to constraint the free functions of the theory and determine conservation laws for the gravitational field equations. The second method that we apply for the study of the evolution of the field equations is the stability analysis of equilibrium points. Particularly, we find solutions with $w_{\text{tot}}=-1$, and we study their stability by means of the Center Manifold Theorem. We show this solution is an attractor in the dilatonic frame but it is an intermediate accelerated solution $a \simeq e^{A t^p}, p:=\frac{2}{2+l}, \quad \frac{32}{57+6 \omega_0}<p<\frac{2}{3}, \;\text{as}\; t\rightarrow \infty$, and not de Sitter solution. The exponent $p$ is reduced, in a particular case, to the exponent already found for the Jordan's and Einstein's frames by \href{http://iopscience.iop.org/article/10.1088/1475-7516/2016/02/027}{A. Cid, G. Leon and Y. Leyva, JCAP 1602, no. 02, 027 (2016)}. We obtain some equilibrium points which represent stiff solutions. Additionally, we find solutions that can be a phantom solution, a solution with $w_{\text{tot}}=-1$ or a quintessence solution. Other equilibrium points mimic a standard dark matter source ($0<w_{\text{tot}}<1$), radiation ($w_{\text{tot}}=\frac{1}{3}$), among other interesting features. For the dynamical system analysis we develop an extension of the method of $F$-devisers. The new approach relies upon two arbitrary functions $h(\lambda, s)$ and $F(s)$. The main advantage of this procedure allows us to perform a phase-space analysis of the cosmological model without the need for specifying the potentials, revealing the full capabilities of the model.

\end{abstract}

\tableofcontents
\section{Introduction}
Various models have been proposed for the explanation of the results which is
followed from the detailed analysis of the recent cosmological data \cite{Teg,Kowal,Komatsu,suzuki11,Ade15}. The observable late time acceleration
have been attributed to the so-called cosmological fluid Dark energy. The
nature of dark energy is unknown and the theoretical approaches to the
problem can be classified in two categories. The first category: the
context of General Relativity an \textquotedblleft exotic\textquotedblright\
matter source which is introduced and provides the late time acceleration of the
universe \cite{Ar1,Ar2,Ar3,Ar4,Ar5}. On the other hand, the second
category: the expansion of the universe is attributed to terms which
follow from the modification of General Relativity (GR), see for instance
\cite{ref1,ref2,ref3,ref4,ref5,ref6,ref7} and references therein. In the
latter theories the new terms, which follow from the modification of the
Einstein-Hilbert action, provide a geometric explanation for the acceleration
of the universe.

{Scalar fields have played an important role on the Dark energy
problem during the last decades. The main characteristic of scalar fields
is that they provide new dynamical variables in the gravitational field
equations, which can modify the evolution of the field equations 
to describe the observations. In addition, scalar fields provide the main
mechanism to explain the early accelerated era of the universe known as
inflation \cite{guth}. Quintessence is one of the most simplest scalar field
model \cite{Ar1} which has many interesting applications in the literature
\cite{qqq01,qqq02,qqq03}. Phantom scalar field is an extension of the
quintessence model, where the scalar field has a non-canonical negative
kinetic energy term. The main characteristic of the phantom field is that
the equation of state parameter for the cosmological fluid is below the
boundary of the cosmological constant ($w=-1$). However, there exist ghosts in that specific model \cite{lurena}. Quintom
scalar field models consist in two scalar fields: a quintessence field and a
phantom field. The main property of the quintom models is that in general,
the two scalar fields are minimally coupled and they can interact in the
kinetic part (and in some models are proposed interactions through the potential function). In the quintom model, the dark energy
equation of state parameter can cross the phantom divide line without the
presence of ghosts  \cite{quin00,Guo:2004fq,Feng:2004ff,Wei:2005nw,Zhang:2005eg,Zhang:2005kj,Lazkoz:2006pa,Lazkoz:2007mx,Setare:2008pz,Setare:2008pc,Leon:2008aq,Leon:2012vt,Leon:2014bta,Leon:2018lnd,Mishra:2018dzq,Marciu:2019cpb,Marciu:2020vve,Dimakis:2020tzc}.} 

{Another two scalar field cosmological model, which has played a
significant role in the description of the early-time acceleration phase of
the universe, is the Chiral cosmology. The theory is defined in the
Einstein frame where the two kinetic energy of the scalar fields defines a
two-dimensional manifold of constant nonzero curvature \cite{atr6,atr7}.
While, the equation of state parameter of the dark energy fluid
in Chiral cosmology is bounded by the values one and minus one. In a recent
study \cite{ndch} it was found that under a quantum transition during the early
universe the equation of state parameter can cross the phantom divide line
and be unbounded and vice versa. In particular, it can
be seen as a model to describe all the components of the dark sector of the
universe, i.e. the dark energy and the dark matter \cite{anchiral1}. It is
clear that multi-scalar field models can provide a simple mechanism,
generated by the variation of an Action Integral, such that to describe
various epochs of the cosmological history \cite%
{Paliathanasis:2018vru,Elizalde:2004mq,Elizalde:2008yf}.}

In the context of this work, we are interested on the Brans-Dicke
gravitational action  in cosmological studies: the so-called Brans-Dicke theory (BDT). Brans and Dicke in 1961
proposed a gravitational action which satisfies Mach's principle \cite%
{bdpaper}, as the prototype of Scalar-tensor theories (STT).  {In general, Scalar-tensor theories (STT) of gravity
\cite{bdpaper,Wagoner:1970vr,OHanlon:1972sdp,OHanlon:1972ysn,Bekenstein:1977rb,Bergmann:1968ve,Nordtvedt:1970uv}
are supported by fundamental physical theories like superstring
theory \cite{Green:1996bh}. The scalar field of BDT says that $\Phi,$
acts as the source for the gravitational coupling with a varying
Newtonian ``constant''  $G\sim \Phi^{-1}.$  This theory satisfies several observational tests including Solar
System tests \cite{Abramovici:1992ah} and Big-Bang nucleosynthesis
constraints \cite{Serna:2002fj,Serna:1995tr}. In extension, the BD parameter $\omega(\Phi),$ can vary depending on the scalar field. 
 Additionally,  a non-zero self-interaction potential $V(\Phi),$ can be incorporated. 
Also, the resulting theory survives astrophysical tests
\cite{Will:1993ns,Will:2005va,Barrow:1996kc}.}
{On the other hand, some ``Extended'' inflationary  models can be found in 
\cite{Barrow:1990nv,Faraoni:2006ik,Liddle:1991am,La:1989za} where the BDT is used 
as the correct
theory of gravity. In this case, the vacuum energy leads
directly to a powerlaw solution \cite{Mathiazhagan:1984vi}, while
the exponential expansion can be obtained when  a cosmological
constant is inserted explicitly into the field equations
\cite{Barrow:1990nv,Kolitch:1994kr,Romero:1992ci}.}

{The action for a general class of STT, can be written in the so-called
Einstein frame (EF) as \cite{Kaloper:1997sh}:
\begin{align}&S_{EF}=\int_{M_4} d{ }^4 x \sqrt{|g|}\left\{\frac{1}{2} R-\frac{1}{2}(\nabla\phi)^2-V(\phi)+\Phi(\phi)^{-2}
\mathcal{L}_{matter}(\mu,\nabla\mu,\Phi(\phi)^{-1}g_{\alpha\beta})\right\},\label{eq1}
\end{align}
where $R$ is the curvature scalar, $\phi$ is the
scalar field, related via conformal transformations with the (Brans-Dicke)
 field $\Phi$,  $(\nabla\phi)^2$ denotes $g^{\mu
\nu}\nabla_\mu\phi\nabla_\nu\phi$, where denotes $\nabla_\alpha$
is the covariant derivative; $V(\phi)$ is the quintessence self-interaction potential,
$\Phi(\phi)^{-2}$ is the coupling function, $\mathcal{L}_{matter}$
is the matter Lagrangian, $\mu$ is a collective name for the
matter degrees of freedom. The energy-momentum matter tensor is defined by
\begin{equation}T_{\alpha
\beta}=-\frac{2}{\sqrt{|g|}}\frac{\delta}{\delta g^{\alpha
\beta}}\left\{\sqrt{|g|}
 \Phi^{-2}\mathcal{L}(\mu,\nabla\mu,\Phi^{-1}g_{\alpha
 \beta})\right\}.\label{Tab}\end{equation}
}
{The issue of regularity of conformal transformations, or the equivalence
of Einstein and Jordan frames have been studied in detail in 
\cite{Magnano:1990qu,Cotsakis:1993vm,Teyssandier:1995wr,Schmidt:1995ws,Cotsakis:1995wt,Capozziello:1996xg,Magnano:1993bd,Faraoni:1998qx,Faraoni:2007yn,Faraoni:2006fx}
and references therein. Using the conformal transformation $\bar{g}_{\alpha
\beta}=\Phi(\phi)^{-1}$ ``constant'' $\omega(\Phi)$ in such way that \newline
$d\phi=\pm \sqrt{\omega(\Phi)+3/2}\Phi^{-1} d\Phi$ and redefining
$\bar{V}(\Phi)=\Phi^2 V(\phi(\Phi))$ the action (\ref{eq1}) can be
written in the Jordan frame (JF) as (see \cite{Coley:2003mj}):
\begin{align}& S_{JF}=\int_{M_4} d{ }^4 x \sqrt{|\bar{g}|}\left\{\frac{1}{2}\Phi \bar{R}-\frac{1}{2}\frac{\omega(\Phi)}{\Phi}(\bar{\nabla}\Phi)^2-\bar{V}(\Phi)
+\mathcal{L}_{matter}(\mu,\nabla\mu,\bar{g}_{\alpha
\beta})\right\},\label{eq1JF}
\end{align}
where a bar is used to denote geometrical objects defined with
respect to the metric $\bar{g}_{\alpha \beta}.$}
{In the STT given by (\ref{eq1JF}), the energy-momentum of the
matter fields is separately conserved. That is
$$\bar{\nabla}^\alpha \bar{T}_{\alpha \beta}=0,$$ where $$\bar{T}_{\alpha
\beta}=-\frac{2}{\sqrt{|\bar{g}|}}\frac{\delta}{\delta
\bar{g}^{\alpha \beta}}\left\{\sqrt{|\bar{g}|}
\mathcal{L}(\mu,\nabla\mu,\bar{g}_{\alpha
 \beta})\right\}.$$ However, when is written in
 EF (\ref{eq1}), this one is no longer the case (although the
overall energy density is conserved). In fact in the EF we find
that
$$Q_\beta\equiv\nabla^\alpha T_{\alpha \beta}=-\frac{1}{2}T\frac{1}{\Phi(\phi)}\frac{\mathrm{d}\Phi(\phi)}{\mathrm{d}\phi}\nabla_{\beta}\phi,\;
 T=T^\alpha_\alpha.$$}
{By making use of the above ``formal'' conformal equivalence between
the Einstein and Jordan frames we can find, for example, that the
theory formulated in the EF with the coupling function
$\Phi(\phi)=\Phi_0 \exp((\phi-\phi_0)/\varpi),\; \varpi\equiv
\pm\sqrt{\omega_0+3/2}$ and potential $V(\phi)=\beta
\exp({(\alpha-2){\varpi}/(\phi-\phi_0)})$ corresponds to the
Brans-Dicke theory (BDT) with a powerlaw potential, i.e.,
$\omega(\Phi)=\omega_0,\; \bar{V}(\Phi)=\beta \Phi^\alpha.$ Exact
solutions with exponential couplings and exponential potentials
(in the EF) were investigated in \cite{Gonzalez:2006cj}.
It was found (see \cite{Coley:2003mj} and references therein) that
typically at early times ($t\rightarrow 0$) the BDT solutions are
approximated by the vacuum solutions and at late times
($t\rightarrow\infty$) by matter dominated solutions, in which the
matter is dominated by the BD scalar field (denoted by $\Phi$ in
the Jordan frame). Exact perfect fluid solutions in STT of gravity
with a non-constant BD parameter $\omega(\Phi)$ have been obtained
by various authors \cite{Mimoso:1994wn,Mimoso:1995ge,Coley:1999yq}.}
Summarizing, in BDT a new degree of freedom is introduced which is
attribute to a scalar field which is nonminimally coupled to gravity. The
importance of that theory is that it is equivalent under conformal
transformation with GR which includes minimally coupled scalar field.
Furthermore, other higher-order theories can be written in terms of
Brans-Dicke field by using Lagrangian multipliers \cite{sotirioureview}. In
the cosmological scenario of a spatially flat Friedmann-Lema{\^{\i}}%
tre-Robertson-Walker geometry we assume the existence of a second perfect
fluid which is described by a scalar-field minimally coupled to gravity. In
this consideration and in the Einstein frame the gravitational field
equation is that of GR in the so-called $\sigma $-models. That is, two
scalar fields with interactions in the kinetic and in the dynamical parts of
the Lagrangian. Integrable cosmological models with non-minimal coupling
have been studied, e.g., in \cite{Kamenshchik:2013dga}. In \cite%
{Kamenshchik:2015cla} it was shown that sometimes is easier to prove
the integrability of the model with non-minimal coupling then the
corresponding model in the Einstein frame. Bianchi I model with non-minimal
coupling has a general solution in the analytic form, but in the case of
zero potential \cite{Kamenshchik:2017ojc}.

In this paper we propose a modified Brans-Dicke theory where the Brans-Dicke
field $\Phi $ is driven by a potential $U(\Phi )$ and the matter content is
modeled by a second scalar field $\psi $ with potential $W(\psi )$. {%
In particular, we define a two-scalar field model in the Jordan frame, where
the field }$\psi ${\ is minimally coupled to gravity. The scalar
field potentials are not specified from the starting point. }So, in order to
specify the unknown potentials, we first express the action in the dilatonic
frame by introducing the dilaton field $\varphi $ with potential $V\left(
\varphi \right) $. The potentials can be derived by applying the method of
group invariant transformations. The existence of a symmetry vector is
important since the latter can be used as an invariant surface to be
defined in the phase-space of the dynamical system. More details on the
application of group invariant transformations in cosmological studies can
be found in \cite{anGRG,antwof,sym1,sym2} and references therein. {%
While a recent application of the point transformations in a two-scalar
field model in the Jordan frame performed in \cite{angrg1}.} 
{The same class of models have been presented in \cite{Cid:2015pja}, for an specific choice of the potential functions.
In this paper we study a more general model and recover previous results of   \cite{Cid:2015pja}.
Particularly, we find solutions with $w_{\text{tot}}=-1$, and we study their stability by means of the Center Manifold Theorem. We show the solution is an attractor in the dilatonic frame but it is an intermediate accelerated solution $a \simeq e^{A t^p}, p:=\frac{2}{2+l}, \quad \frac{32}{57+6 \omega_0}<p<\frac{2}{3}, \;\text{as}\; t\rightarrow \infty$, and not a de Sitter solution. The exponent $p$ is reduced, in a particular case, to the exponent already found for the Jordan's and Einstein's frames by \cite{Cid:2015pja}.}

{The present analysis is fairly more general because we consider the potentials to be free functions and then
find the generic features of the dynamical system, under the assumption that the system can be written in a closed form.} In this regard, we propose a general method for the construction of the phase space that  relies in the specification of two arbitrary functions $F(s)$ and $h(s,\lambda)$. The equilibrium points with
$s$ constant such that $h$ is only a function of $\lambda$ (depending on the choice of $W$), and with  $F$ identically zero. They are easily found due to  the problem can
be reduced in one dimension. When $F(s)$ is not trivial, we discuss a general classification that can be implemented straightforwardly for any of the specific
choices of $F$ for the scalar field potentials commonly used in the literature. On the other hand, the search of the equilibrium points with $\lambda\neq 0$, is not an easy
task, and the success on it depends crucially on the choice of $h(s,\lambda)$.

The main advantage of this procedure is that it allows us to perform a
phase-space analysis of the cosmological model, without the need for
specifying potentials. {Recent reviews on the phase space
analysis of various cosmological models can be found in \cite{dynrev,Quiros:2019ktw}.} This
phase-space and stability examination let us to bypass the non-linearities
and complications of the cosmological equations, which prevent complete
analytical treatments by obtaining a qualitative description of the global
dynamics of these scenarios, which is independent of the initial conditions
and the specific evolution of the universe. Furthermore, we are able to calculate various
observable quantities in these asymptotic solutions, such as
the dark-energy and total equation-of-state parameters, the deceleration
parameter, the various density parameters, etc. However, in order to remain in
general, we extend beyond the usual procedure \cite%
{Burd:1988ss,REZA,Copeland:1997et,Coley:2003mj,Gong:2006sp,
Setare:2008sf,Chen:2008ft,Gupta:2009kk,Matos:2009hf,
Copeland:2009be,Leyva:2009zz,
Farajollahi:2011ym,UrenaLopez:2011ur,Escobar:2011cz,
Escobar:2012cq,Xu:2012jf,Leon:2013qh}. As far as we know this methodology
has not been introduced  in the literature yet; although, it is inspired in the
method of the $F$- devisers extensively used in the relativistic setting in
\cite{Rendall:2004ic,Fang:2008fw,Leon:2008de,Matos:2009hf,Leyva:2009zz,
UrenaLopez:2011ur} and that has been formalized in \cite%
{Escobar:2013js,Fadragas:2013ina,delCampo:2013vka}.  {Thefereore, this method is a generalization of the procedure  presented for single scalar field models.}

For illustrating the advantages of the method we consider some specific forms of the potentials $%
V\left( \varphi \right) $ and $W\left( \psi \right) $ which lead to
specific forms on functions $F(s)$ and $h(s,\lambda )$. For the
Brans-Dicke field $\Phi ~$we consider a power-law potential, where in terms
of the field $\varphi $ has the exponential form $V(\varphi
)=V_{0}e^{l\varphi }$. As far as, we study
the cases of the second scalar field  where the potential is (a) exponential and (b) power-law.
Finally, we comment about general features of the equilibrium points of  the dynamical system.

Comparing with the Jordan-Brans-Dicke theory introduced and studied in  \cite{Cid:2015pja} . We have in the Jordan's and Einstein's frames the following.
In the Jordan frame the potentials of \cite{Cid:2015pja} are (we have renamed the original constants as $\lambda_U$ and $\lambda_W$): 
$$U(\Phi)=U_0 \Phi^{2-\lambda_U \sqrt{\omega_0+\frac{3}{2}}}, \; W(\psi)=W_0 e^{-\lambda_W \psi}.$$
Therefore, the fields of this theory in the dilatonic action will be the dilaton $\varphi$ with potential $$V(\varphi)= U_0 e^{\left(1-\lambda_U \sqrt{\omega_0+\frac{3}{2}}\right)\varphi},$$ and a second scalar field $\psi$ with potential $$W(\psi)=W_0 e^{-\lambda_W \psi}.$$
Hence, the model studied in \cite{Cid:2015pja} can be considered as a special case of the model studied in section \ref{sect:4.1} Case: $W(\protect\psi)=W_0 e^{k \protect\psi}$ and $V(\protect%
\varphi)=V_0 e^{l \protect\varphi}$ with $k=-\lambda_W, l=1-\lambda_U \sqrt{\omega_0+\frac{3}{2}}, \omega_0>-\frac{3}{2}$.

The paper is organized as follows. Our model is defined in Section \ref{Sect:2}. The point-like Lagrangian and some
exact solutions by using group invariant transformations are presented in
Section \ref{Sect:3}. In Section \ref{Sect:4}, we rewrite the field equations in dimensionless
variables and we end up with a five first-order differential-algebraic system
with two unknown functions which are related to the potentials of the two
scalar fields. We study the evolution of the field equations
by using dynamical system tools for some explicitly forms of the potentials. In particular, we consider the cases where the Brans-Dicke scalar
field is power law while the minimally coupled field has an exponential,
potential or a power law potential. The case: $W(\protect\psi)=W_0 e^{k \protect\psi}$ and $V(\protect%
\varphi)=V_0 e^{l \protect\varphi}$ is studied in Section \ref{sect:4.1}, whereas, the case: $W(\protect\psi)=W_0 \protect\psi^{k}$ and $V(\protect%
\varphi)=V_0 e^{l \protect\varphi}$ is studied in Section \ref{sect:4.2}. Going to the general set up, we find generic features of the dynamics without specifying the potentials in Section \ref{Sect:5}. This allows us to find generic results that are independent of the model choice. The cosmological implications of the model at hand are discussed in Section \ref{Sect:6}. Finally, our conclusions and discussions
are given in Section \ref{Sect:7}.

\section{Gravitational model}
\label{Sect:2}
Considering the gravitational Action integral to be
\begin{align}
&S=\int d^{4}x\sqrt{-g}\left\{ \frac{\Phi R}{2}-\frac{\omega _{0}}{2\Phi }%
\nabla ^{\mu }\Phi \nabla _{\mu }\Phi -U(\Phi )  -\frac{1}{2}\nabla ^{\mu
}\psi \nabla _{\mu }\psi -W(\psi )\right\} ,  \label{action}
\end{align}%
 where $\Phi $ is the Brans-Dicke field and $\psi $ representing a
quintessence field. $U(\Phi )$ and $W(\psi )$ are the corresponding
potentials for the scalar fields. For the sake of simplicity and without loss of generality we rescale the Brans-Dicke field $\Phi $ and the
associated potential $U(\Phi )$ as,
\begin{equation}
\Phi =e^{\varphi }\quad \quad \text{and}\quad \quad U(\Phi )=e^{\varphi
}V(\varphi ).
\end{equation}

Consequently, under a conformal transformation the action (\ref{action}) is
transformed into the dilatonic action:
\begin{align}
& S=\int d^{4}x\sqrt{-g}e^{\varphi }\left\{ \frac{R}{2}-\frac{\omega _{0}}{2}%
\nabla ^{\mu }\varphi \nabla _{\mu }\varphi -V(\varphi ) -\frac{1}{2}%
e^{-\varphi }\nabla ^{\mu }\psi \nabla _{\mu }\psi -e^{-\varphi }W(\psi
)\right\}. \label{dilatonic}
\end{align}%
The field equations associated to action (\ref{dilatonic}) are given by:
\begin{subequations}
\begin{align}
& G_{\mu \nu }=(1+\omega _{0})\left( \nabla _{\mu }\varphi \nabla _{\nu
}\varphi -\frac{1}{2}g_{\mu \nu }\nabla _{\alpha }\varphi \nabla ^{\alpha
}\varphi \right)   -g_{\mu \nu }\left( \frac{1}{2}\nabla _{\alpha }\varphi \nabla ^{\alpha
}\varphi +V(\varphi )\right) \notag \\
& +\nabla _{\mu }\nabla _{\nu }\varphi -g_{\mu
\nu }\nabla ^{2}\varphi +e^{-\varphi }T_{\mu \nu }^{(\psi )}, \\
& \nabla ^{2}\varphi +\nabla _{\mu }\varphi \nabla ^{\mu }\varphi =\frac{2}{%
3+2\omega _{0}}(V^{\prime }(\varphi )-V(\varphi )) +\frac{e^{-\varphi }}{%
3+2\omega _{0}}T^{(\psi )}, \\
& \nabla ^{2}\psi =W^{\prime }(\psi ),
\end{align}%
\end{subequations}
where $\nabla ^{2}\equiv \nabla ^{\mu }\nabla _{\mu }$ and $T^{(\psi )}=-\nabla ^{\mu }\psi \nabla _{\mu }\psi -4W(\psi )$ is
the trace of the energy-momentum tensor $T_{\mu \nu }^{(\psi )}=\nabla _{\mu }\psi \nabla _{\nu }\psi -g_{\mu \nu
}\left( \frac{1}{2}\nabla ^{\alpha }\psi \nabla _{\alpha }\psi +W(\psi
)\right)$.

We assume the geometry which describes the universe is that of
spatially flat Friedmann-Lema{\^{\i}}tre-Robertson-Walker spacetime
\begin{equation}  \label{FRW_jordan}
ds^{2}=-dt^{2}+a^{2}(t)\delta _{ij}dx^{i}dx^{j}.
\end{equation}

For the latter line element and for the comoving observer $\left(
u^{a}=\delta _{t}^{a}~,~u^{a}u_{a}=-1\right)$, we calculate the field
equations to be

\begin{subequations}
\label{FE_jordan}
\begin{align}
& 3H^{2}=\frac{\omega _{0}}{2}\dot{\varphi}^{2}-3H\dot{\varphi}+V(\varphi
)+e^{-\varphi }\left( \frac{1}{2}\dot{\psi}^{2}+W(\psi )\right),
\label{fe.01} \\
& \dot{H}=-\frac{1}{2}\omega _{0}\dot{\varphi}^{2}+2H\dot{\varphi}+\frac{%
V^{\prime }(\varphi )-V(\varphi )}{3+2\omega _{0}} -\frac{e^{-\varphi
}(1+\omega _{0})\dot{\psi}^{2}}{3+2\omega _{0}}-\frac{2e^{-\varphi }W(\psi )%
}{3+2\omega _{0}},  \label{fe.02} \\
& \ddot{\varphi}+3H\dot{\varphi}+\dot{\varphi}^{2}=2\frac{V(\varphi
)-V^{\prime }(\varphi )}{3+2\omega _{0}} -e^{-\varphi }\frac{\dot{\psi}%
^{2}-4W(\psi )}{3+2\omega _{0}},  \label{fe.03} \\
& \ddot{\psi}+3H\dot{\psi}+W^{\prime }(\psi )=0. \label{fe.04}
\end{align}%
where (\ref{fe.01}) is the modified first Friedmann's equation, equation (%
\ref{fe.02}) is the Raychaudhuri (acceleration) equation and equations (\ref{fe.03}), (\ref%
{fe.04}) are the \textquotedblleft Klein-Gordon\textquotedblright\ equations
in which the two scalar fields should satisfy.

In the following section, we determine the point-like Lagrangian for the
field equations as we search for solutions by using\ the method of
group invariant transformations.

\section{Minisuperspace approach and exact solutions}
\label{Sect:3}
From the Action integral (\ref{dilatonic}) and for the FRW spacetime with
line element

\end{subequations}
\begin{equation}
ds^{2}=-N^{2}\left( t\right) dt^{2}+a^{2}\left( t\right) \left(
dx^{2}+dy^{2}+dz^{2}\right) ,  \label{mn.01}
\end{equation}%
the following Lagrangian density can be defined by
\begin{align}
&\mathcal{L}=\frac{1}{N}\left( -3e^{\varphi }a\dot{a}^{2}-3e^{\varphi }a^{2}%
\dot{a}\dot{\varphi}+\frac{\omega _{0}}{2}a^{3}e^{\varphi }\dot{\varphi}^{2}+%
\frac{1}{2}a^{3}\dot{\psi}^{2}\right)  -Na^{3}\left( e^{\varphi }V(\varphi
)+W\left( \psi \right) \right) ,  \label{mn.02}
\end{align}%
 where the field equations (\ref{fe.01})-(\ref{fe.04}) follow from the
Euler-Lagrange with respect to the variables $\left\{ N,a,\varphi ,\psi
\right\} $. Lagrangian (\ref{mn.02}) describes a singular system of
second-order differential equations, because the determinant of the Hessian
matrix is zero, i.e. $\left\vert \frac{\partial \mathcal{L}}{\partial \dot{x}%
^{i}\partial \dot{x}^{j}}\right\vert =0$. Specifically, the field equations
form a constraint dynamical system \cite{dirac}, with constraint equation $%
\frac{\partial \mathcal{L}}{\partial N}=0.$

Without loss of generality we can consider that $N\left( t\right) =N\left(
a\left( t\right) ,\varphi \left( t\right) \right) $, where now Lagrangian (%
\ref{mn.02}) is autonomous and admits the symmetry vector field $\partial
_{t}$, where the corresponding conservation law is the Hamiltonian function $%
\mathcal{H}=const$. However, from the first modified Friedmann's equation we
have that $\mathcal{H}=0$.

We consider that $N=\bar{N}\left( t\right) e^{-\varphi /2}$ and $%
a=Ae^{-\varphi /2}$, where now the line element (\ref{mn.01}) becomes%
\begin{equation}
ds^{2}=e^{-\varphi }\left( -\bar{N}^{2}\left( t\right) dt^{2}+A^{2}\left(
t\right) \left( dx^{2}+dy^{2}+dz^{2}\right) \right) ,  \label{mn.03}
\end{equation}%
while the Lagrangian of the field equations is written as follows
\begin{align}
&\mathcal{L}=\frac{1}{\bar{N}}\left[ -3A\dot{A}^{2}+\frac{1}{2}A^{3}\left(
\Omega _{0}\dot{\varphi}^{2}+e^{-\varphi }\dot{\psi}^{2}\right) \right] -%
\bar{N}A^{3}\left( e^{-\varphi }V(\varphi )+e^{-2\varphi }W\left( \psi
\right) \right) ,  \label{mn.04}
\end{align}%
 in which $\Omega _{0}=\frac{3+2\omega _{0}}{2}.$

Lagrangian (\ref{mn.04}) is nothing else that the cosmological model of two
scalar fields minimally coupled in gravity but with interaction in the
kinetic and dynamic terms. Specifically, Lagrangian (\ref{mn.04}) describes
the field equations for the action integral
\begin{align}
&\bar{S}=\int d^{4}x\sqrt{-\bar{g}}\left( \bar{R}\left( \bar{g}_{\mu \nu
}\right) -\frac{\Omega _{0}}{2}\bar{g}^{\mu \nu }\varphi _{;\mu }\varphi
_{;\nu } -\frac{1}{2}e^{-\varphi }\bar{g}^{\mu \nu }\psi _{;\mu }\psi _{;\nu
}-e^{-\varphi }V\left( \varphi \right) -e^{-2\varphi }W\left( \psi \right)
\right),  \label{mn.05}
\end{align}%
 where $\bar{g}_{\mu \nu }=e^{\varphi }g_{\mu \nu }$.

The last action belong to the action of the so-called nonlinear $\sigma $%
-models \cite{sigma1}. On the other hand, the action integral (\ref{mn.05})
can be seen like a complex scalar field where the norm of the complex
plane is not defined by the unitary matrix but from a space of constant
curvature $E_{B}^{A}=$\textrm{diag}$\left( \Omega _{0},e^{-\varphi }\right) $%
, with Ricciscalar $R_{\left( 2\right) }=-\frac{1}{2\Omega _{0}}$. Finally,
because of the constraint equation any solution of the dynamical system with
Lagrangian (\ref{mn.04}) will be also a solution for the system (\ref{mn.02}%
) (for a discussion see \cite{anGRG}). Some exact solutions for cosmological
models of the form of (\ref{mn.05}) can be found in \cite{antwof,sigma2} and
references therein. In the following without loss of generality in (\ref%
{mn.04}) we select $\bar{N}=1$.

In order to specify the unknown potentials $V\left( \varphi \right) $ and $%
W\left( \psi \right) $ we apply the method of group invariant
transformations. We find that for
\begin{equation}
V\left( \varphi \right) =V_{0}e^{\left( \beta -1\right) \varphi }~\
,~W\left( \psi \right) =W_{0}\psi ^{2\beta }  \label{mn.06}
\end{equation}%
Lagrangian (\ref{mn.04}) admits the Noether point symmetry vector%
\begin{equation}
X=\left( 2-\beta \right) t\partial _{t}+\frac{\left( 2-\beta \right) }{3}%
a\partial _{t}+2\partial _{\varphi }+\psi \partial _{\psi },  \label{mn.07}
\end{equation}%
where the corresponding conservation law is\footnote{%
The constraint equation $\frac{\partial L}{\partial \bar{N}}=0$ have been
applied.}
\begin{equation}
I_{X}=\left( \beta -2\right) a^{2}\dot{a}+\Omega _{0}a^{3}\dot{\varphi}+%
\frac{1}{2}e^{-\varphi }a^{3}\psi \dot{\psi}.  \label{mn.08}
\end{equation}

Considering now that $\beta =2$, and the value of the conservation law is
zero, that is, $I_{X}=0$, then from (\ref{mn.08}) follows%
\begin{equation}
\dot{\varphi}=-\frac{1}{2\Omega _{0}}e^{-\varphi }\psi \dot{\psi}\rightarrow
e^{\varphi }=-\frac{1}{4\Omega _{0}}\psi ^{2}+c.
\end{equation}%
or%
\begin{equation*}
\varphi =\ln \left( -\frac{1}{4\Omega _{0}}\psi ^{2}+c\right)
\end{equation*}%
By replacing in the Hamiltonian function we have{\small
\begin{align}
&\mathcal{H}=-3A\dot{A}^{2}+\frac{1}{2}A^{3}\left( \frac{1}{4\Omega _{0}}%
\frac{\left( 1-\frac{1}{4\Omega _{0}}\right) \psi ^{2}+c}{\left( -\frac{1}{%
4\Omega _{0}}\psi ^{2}+c\right) ^{2}}\right) \dot{\psi}^{2} +A^{3}\left(
V_{0}+W_{0}\frac{\psi ^{4}}{\left( -\frac{1}{4\Omega _{0}}\psi ^{2}+c\right)
^{2}}\right) =0  \label{mn.09}
\end{align}%
} where in the limit $c=0$, the field equations correspond to that of GR
with a cosmological constant and a stiff matter. The latter follows from the
kinetic part of the scalar field $\dot{\Psi},$ where
\begin{equation}
d\Psi =\sqrt{\frac{\left( 1-\frac{1}{4\Omega _{0}}\right) \psi ^{2}+c}{%
\left( -\frac{1}{4\Omega _{0}}\psi ^{2}+c\right) ^{2}}}d\psi .  \label{mn.10}
\end{equation}%
For a nonzero constant $c$, (\ref{mn.09}) corresponds to the first
Friedmann's equation of GR with a minimally coupled scalar field where the
general solution is given in \cite{an3}. In the limit where $\Omega _{0}=-%
\frac{1}{4}$, i.e. $\omega _{0}=-\frac{5}{4}$, from (\ref{mn.10}) we have
the closed-form expression $\psi =\sqrt{c}\tanh \Psi $, where (\ref{mn.09})
becomes%
\begin{equation}
\mathcal{H}=-3A\dot{A}^{2}+\frac{1}{2}A^{3}\dot{\Psi}^{2}+A^{3}\sinh
^{4}\left( \Psi \right) =0
\end{equation}%
that is, of a quintessence field with the hyperbolic potential $W\left(
\Psi \right) =\sinh ^{4}\left( \Psi \right) $. \

In general, for $\beta \neq 2$ and from the symmetry vector (\ref{mn.07}) we
define the Lagrange system%
\begin{equation}
\frac{dt}{\left( 2-\beta \right) t}= \frac{da}{\frac{\left( 2-\beta \right)
}{3}a}=\frac{d\varphi }{2}=\frac{d\psi }{\psi }
\end{equation}%
from where we define the invariants $u=At^{\frac{1}{3}},~v=e^{\varphi }t^{%
\frac{2}{\beta -2}},~w=\psi t^{^{\frac{1}{\beta -2}}}\,$. Recalling that a
Noether symmetry is also a Lie point symmetry for the field equations.

The invariants can be used to reduce the order of the differential equations
or to determine a special solution. \ Considering that the invariants are
constants, i.e. $\left( u,v,w\right) \rightarrow \left( A_{0},e^{\varphi
_{0}},\psi _{0}\right) $, then, we observe that
\begin{equation}
A\left( t\right) =A_{0}t^{-\frac{1}{3}},~e^{\varphi }=e^{\varphi _{0}}t^{-%
\frac{2}{\beta -2}},~\psi =\psi _{0}t^{-\frac{1}{\beta -2}}  \label{mn.11}
\end{equation}%
solve the field equations for the gravitational field equations with
Lagrangian (\ref{mn.04}) and $\bar{N}\left( t\right) =1$, for the potentials
(\ref{mn.06}) when the constants $W_{0},V_{0},~\Omega _{0}$ and $\beta $ are
related as follows

\begin{align}
& W_{0}=\frac{2\beta -5}{2\beta \left( \beta -2\right) ^{2}}e^{\varphi
_{0}}\left( \psi _{0}\right) ^{2\left( 1-\beta \right) }~,\\
& V_{0}=e^{\left(
1-\beta \right) \varphi _{0}}\left( \frac{5\left( \psi _{0}\right)
^{2}+8e^{\varphi _{0}}\beta \omega _{0}}{2\beta \left( \beta -2\right) ^{2}}%
\right),
\end{align}%
and~%
\begin{equation}
\Omega _{0}=-\frac{e^{-\varphi _{0}}}{12}\left( 2e^{2\varphi _{0}}\beta
^{2}-8e^{\varphi _{0}}\beta +3\psi _{0}^{2}-8e^{\varphi _{0}}\right)
\end{equation}

Solution (\ref{mn.11}) is a special solution of the field equations in the
Einstein frame. Going back now in the Jordan frame, where
\begin{equation}
a\left( \tau \right) =A\left( \tau \right) e^{-\varphi \left( \tau \right)
/2}~,~e^{-\varphi /2}dt=d\tau
\end{equation}%
we have~~$t=\sigma _{1}\left( \varphi _{0},\beta \right) t^{\frac{\beta -2}{%
\beta -1}},~\beta \neq 1,2$, and $t=\sigma _{2}\left( \varphi _{0}\right)
e^{t}~ $for$~\beta =1$, hence~for the scale-factor holds $a\left( \tau
\right) \simeq \tau ^{\frac{5-\beta }{3\left( \beta -1\right) }}~,~\beta
\neq 1,2,~$and $a\left( \tau \right) \simeq e^{\sigma _{2}\tau }$. The
latter is a de Sitter solution while the first one is a perfect fluid
solution in which
\begin{equation*}
w_{\text{eff}}=\frac{p_{\text{eff}}}{\rho _{\text{eff}}}=-\frac{3\beta -7}{%
\beta -5},
\end{equation*}%
where there exists acceleration, i.e. $w_{\text{eff}}<-\frac{1}{3}$,\\ for $%
\beta \in \left( -\infty ,2\right) \cup \left( 5,+\infty \right) $, while
for $\beta =\frac{13}{5}$, we have a radiation solution and for $\beta =%
\frac{7}{3}$ the solution is that of a pressureless fluid.

We continue our analysis with the equilibrium point analysis for the gravitational
field equations, but we keep now the  potentials unspecified.

\section{The dynamical system}
\label{Sect:4}

We define the normalized variables in order to express the above equations as an autonomous closed dynamical
system 
\begin{equation}
\label{coordinates-x-y-z}
x=\frac{\dot{\varphi}}{\sqrt{6}H},\quad y=\frac{\sqrt{V(\varphi )}}{\sqrt{3}H%
},\quad z=\frac{e^{-\frac{\varphi }{2}}\dot{\psi}}{\sqrt{6}H},
\end{equation}%
and the auxiliary variables 
\begin{align}
& s=-\frac{V^{\prime }(\varphi )}{V(\varphi )},\quad \lambda =-e^{\frac{%
\varphi }{2}}\frac{W^{\prime }(\psi )}{W(\psi )}, \quad \Gamma _{\varphi }=%
\frac{V(\varphi )V^{\prime \prime }(\varphi )}{V^{\prime}(\varphi) ^2},\quad \Gamma
_{\psi }=\frac{W(\psi )W^{\prime \prime }(\psi )}{W^{\prime}(\psi) ^2}.
\end{align}%
which are related by
\begin{equation}
-\sqrt{6}x+x^{2}\omega _{0}+y^{2}+z^{2}+\frac{e^{-\varphi }W(\psi )}{3H^{2}}%
=1.
\end{equation}%
Since by definition $\Gamma _{\varphi }:=\frac{V(\varphi )V^{\prime \prime
}(\varphi )}{V^{\prime 2}}$ depends only on $\varphi $ and simultaneously $%
\varphi $ is an implicit function of $s$ through $s=-\frac{V^{\prime
}(\varphi )}{V(\varphi )}$ it follows $\Gamma _{\varphi }=f(s)$.
Furthermore, since by definition $\lambda :=-e^{\frac{\varphi }{2}}\frac{%
W^{\prime }(\psi )}{W(\psi )}$, i.e., it depends on both $\varphi $ and $%
\psi $, thus, using the implicit relation between $\varphi $ and $s$ through
$s=-\frac{V^{\prime }(\varphi )}{V(\varphi )}$ and between $\psi $ and $%
\Gamma _{\psi }$ through $\Gamma _{\psi }=\frac{W(\psi )W^{\prime \prime
}(\psi )}{(W^{\prime 2}}$, we obtain $\lambda =g(s,\Gamma _{\psi })$. Assume
that $\lambda =g(s,\Gamma _{\psi })$ can be explicitly solved for $\Gamma
_{\psi }$, say $\Gamma _{\psi }=h(s,\lambda )$.
Then, the evolution
equations are
\begin{subequations}
\label{syst1}
\begin{align}
& x^{\prime }=x\left[ \frac{3(s-1)y^{2}}{2\omega _{0}+3}+\frac{18}{2\omega
_{0}+3}+\frac{6\omega _{0}z^{2}}{2\omega _{0}+3}-3\right] +\frac{\sqrt{6}%
(s-1)y^{2}}{2\omega _{0}+3}   +\frac{3x^{3}\omega _{0}(2\omega _{0}+1)}{2\omega _{0}+3} \notag \\
& -\frac{\sqrt{6}%
x^{2}(8\omega _{0}+3)}{2\omega _{0}+3}  +\frac{2\sqrt{6}}{2\omega _{0}+3}-%
\frac{3\sqrt{6}z^{2}}{2\omega _{0}+3}, \end{align}
\begin{align}
& y^{\prime }=-\frac{\sqrt{6}xy(2(s+4)\omega _{0}+3s)}{4\omega _{0}+6}+\frac{%
3(s-1)y^{3}}{2\omega _{0}+3} + \frac{3x^{2}y\omega _{0}(2\omega _{0}+1)}{%
2\omega _{0}+3} \notag \\
& +y\left[ \frac{6}{2\omega _{0}+3}+\frac{6\omega _{0}z^{2}}{2\omega _{0}+3}%
\right], \end{align}
\begin{align}
& z^{\prime }=\sqrt{\frac{3}{2}}\lambda +y^{2}\left[ \frac{3(s-1)z}{2\omega
_{0}+3}-\sqrt{\frac{3}{2}}\lambda \right] +x^{2}\left[ \frac{3\omega
_{0}(2\omega _{0}+1)z}{2\omega _{0}+3}-\sqrt{\frac{3}{2}}\lambda \omega _{0}%
\right] +  \notag \\
& +x\left[ 3\lambda -\frac{\sqrt{6}(10\omega _{0}+3)z}{4\omega _{0}+6}\right]
+\frac{6\omega _{0}z^{3}}{2\omega _{0}+3}  -\sqrt{\frac{3}{2}}\lambda z^{2}+%
\left[ \frac{6}{2\omega _{0}+3}-3\right] z, \end{align}
\begin{align}
& \lambda ^{\prime }=\sqrt{\frac{3}{2}}\lambda \left[ x-2(h(s,\lambda
)-1)\lambda z\right] ,  \label{4.4d} \\
& s^{\prime }=-\sqrt{6}xF(s),  \label{4.4e}
\end{align}%
where $F(s):=s^{2}(f(s)-1)$.

We have a dynamical system for the state vector $(x,y,z, \lambda,s)$ defined
in the phase space
\end{subequations}
\begin{align}
\left\{(x,y,z, \lambda,s)\in \mathbb{R}^5: -\sqrt{6} x + x^2 \omega_0 +y^2 +z^2 \leq 1,
\lambda \in \mathbb{R}, s \in \mathbb{R} \right\},
\end{align}
whose evolution is given respectively by \eqref{syst1}.

Defining the function \begin{equation}
C(x,y,z,s,\lambda)=-x^2 \omega _0+\sqrt{6} x-y^2-z^2+1\geq 0,
\end{equation} and calculating the total derivative we have\\
\begin{equation}
C'=C\Big[\frac{6 \left((s-1) y^2+3 x^2+2 \sqrt{6} x-3 z^2+2\right)}{2 \omega _0+3} +6 x^2 \omega _0-6 x^2  -5 \sqrt{6} x+6
   z^2-\sqrt{6} \lambda  z\Big].
	\end{equation}
From this, it follows that if we take the initial conditions over the surface $C=0$, the solutions remain on this surface all the time. And if we take the initial conditions on the half-space $C>0$, the solutions remain on this region for all the time.
 Estimating 
\begin{equation*}
\frac{6 \left((s-1) y^2+3 x^2+2 \sqrt{6} x-3 z^2+2\right)}{2 \omega _0+3} +6 x^2 \omega _0-6 x^2-5 \sqrt{6} x+6
   z^2-\sqrt{6} \lambda  z,
	\end{equation*}
	we can see how the errors propagate if we take the initial conditions on the surface\\ $C(x,y,z,s,\lambda)=C_0$, with $C_0$ arbitrarily small.

 Explicitly, in order to obtain an autonomous dynamical system . Firstly, it is necessary to determine a specific potential form
$V(\varphi)$ and $W(\psi)$.  However, one could alternatively
handle the potential differentiations when $F$ can be expressed as an explicit one-valued function of $s$,
that is $F=F(s)$, as well as it can be defined an explicit function $h=h(s,\lambda)$ for some examples. Therefore, it results to a closed dynamical
system for $s$, $\lambda$,  and a set of normalized-variables.  A similar approach
has been applied in isotropic (FRW) scenarios
\cite{Rendall:2004ic,Fang:2008fw,Leon:2008de,Matos:2009hf,Leyva:2009zz,
UrenaLopez:2011ur}. However, we will improve it for the purpose 
of the present work . Such a procedure is
possible for general physical potentials, and for the usual ansatzes of
the cosmological literature. It results to very simple forms for $F(s)$, as
can be seen in Table \ref{fsform}.

In order to continue we consider some specific forms of the potentials $%
V\left( \varphi \right) $ and $W\left( \psi \right) $ which lead to
specific forms on the functions $F(s)$ and $h(s,\lambda )$. For the
Brans-Dicke field $\Phi ~$we consider a power-law potential where in terms
of the field $\varphi $ has the exponential form $V(\varphi
)=V_{0}e^{l\varphi }$. As far as it concerns in the second scalar field we study
the cases where the potential is (a) exponential and (b) power-law.
Finally, we comment about general features of the equilibrium points of %
\eqref{syst1} for arbitrary $h(\lambda, s)$ and $F(s)$ functions.

\begin{table}
\centering
\caption{The function $F(s)$ for the most common quintessence potentials
\protect\cite{Escobar:2013js}.}
\label{fsform}
\begin{tabular*}{\columnwidth}{@{\extracolsep{\fill}}lll@{}}
\hline
\multicolumn{1}{@{}l}{Potential} & References & $F(s)$\\
\hline
$V(\varphi)=V_{0}e^{-k\varphi}+V_1$ & \cite{Yearsley:1996yg,Pavluchenko:2003ge,Cardenas2003} & $-s(s-k)$ \\
$V(\varphi)=V_{0}\left[e^{\alpha\varphi}+e^{\beta\varphi}\right]$ & \cite{Barreiro:1999zs,Gonzalez:2007hw,Gonzalez:2006cj} & $-(s+\alpha)(s+\beta)$\\
$V(\varphi)=V_{0}\left[\cosh\left( \xi \varphi \right)-1\right]$ & \cite{Ratra:1987rm,Wetterich:1987fm,Matos:2009hf,Copeland:2009be,Leyva:2009zz,Pavluchenko:2003ge,delCampo:2013vka,Sahni:1999qe,Sahni:1999gb,Lidsey:2001nj,Matos:2000ng}
& $-\frac{1}{2}(s^2-\xi^2)$ \\
$V(\varphi)=V_{0}\sinh^{-\alpha}(\beta\varphi)$ & \cite{Ratra:1987rm,Wetterich:1987fm,Copeland:2009be,Leyva:2009zz,Pavluchenko:2003ge,Sahni:1999gb,UrenaLopez:2000aj}
& $\frac{s^2}{\alpha}-\alpha\beta^2$\\
\hline
\end{tabular*}
\end{table}

\subsection{Case: $W(\protect\psi)=W_0 e^{k \protect\psi}$ and $V(\protect%
\varphi)=V_0 e^{l \protect\varphi}$}
\label{sect:4.1} In this example we have $\Gamma_\psi=1, \Gamma_\varphi=1$,
thus $f(s)=1, F(s)=0$ and $h(s,\lambda)=1$. Furthermore, $s=-l=\text{const.}$%
. In this particular, the system \eqref{syst1} simplifies to
\begin{subequations}
\label{syst1exp_exp}
\begin{align}
&x^{\prime }= x \left[-\frac{3 (l+1) y^2}{2 \omega_0+3}+\frac{18}{2
\omega_0+3}+\frac{6 \omega_0 z^2}{2 \omega_0+3}-3\right]  -\frac{\sqrt{6 }
(l+1) y^2}{2 \omega_0+3}  +\frac{3 x^3 \omega_0 (2 \omega_0+1)}{2 \omega_0+3} \notag \\
& -\frac{\sqrt{6} x^2 (8
\omega_0+3)}{2 \omega_0+3} +\frac{2 \sqrt{6}}{2 \omega_0+3}-\frac{3 \sqrt{6}
z^2}{2 \omega_0+3}, \\
& y^{\prime }=\frac{\sqrt{6} x y (2 (l-4) \omega_0+3l)}{4 \omega_0+6}-\frac{%
3 (l+1) y^3}{2 \omega_0+3} +\frac{3 x^2 y \omega_0 (2 \omega_0+1)}{2
\omega_0+3}  \notag \\
& +y \left[\frac{6}{2 \omega_0+3}+\frac{6 \omega_0 z^2}{2 \omega_0+3}\right],
\\
& z^{\prime }=\sqrt{\frac{3}{ 2}} \lambda -y^2 \left[\frac{3 (l+1) z}{2
\omega_0+3}+\sqrt{\frac{3}{2}} \lambda \right] +x^2 \left[\frac{3 \omega_0 (2
\omega_0+1) z}{2 \omega_0+3}-\sqrt{\frac{3}{2}} \lambda \omega_0\right]+
\notag \\
& +x \left[3 \lambda -\frac{\sqrt{6} (10 \omega_0+3) z}{4 \omega_0+6}\right]+%
\frac{6 \omega_0 z^3}{2 \omega_0+3}  -\sqrt{\frac{3}{2}} \lambda z^2+\left[%
\frac{6}{2 \omega_0+3}-3\right] z, \\
& \lambda^{\prime }= \sqrt{\frac{3}{2}} \lambda x.
\end{align}
The system is a form-invariant under the change $(y,
z,\lambda)\rightarrow (-y,-z, -\lambda)$. Therefore, without losing generality we
can investigate just the sector $y\geq 0, z\geq 0, \lambda\geq 0$.
Henceforth, we will focus on the stability properties of the system %
\eqref{syst1exp_exp} for the state vector $(x,y,z, \lambda)$ defined in the
phase space
\end{subequations}
\begin{align}
&\Big\{(x,y,z, \lambda): -\sqrt{6} x + x^2 \omega_0 +y^2 +z^2 \leq 1,  y\geq
0, z\geq 0, \lambda\geq 0 \Big\},
\end{align}
whose evolution is given by \eqref{syst1exp_exp}.

The equilibrium points of the system \eqref{syst1exp_exp} are the following:

\begin{enumerate}
\item[$P_1$:] $(x,y,z,\lambda)=\left(0,\frac{\sqrt{2}}{\sqrt{l+1}},0,0\right)
$.\newline
There exists for $2 \omega _0+3\neq 0, l\geq 1$.\newline
The eigenvalues are \newline
$\left\{0,-3,-\frac{1}{2} \left(3+ \frac{\sqrt{-48 l+18 \omega _0+75}}{\sqrt{%
2 \omega _0+3}}\right),-\frac{1}{2} \left(3-\frac{\sqrt{-48 l+18 \omega _0+75%
}}{\sqrt{2 \omega _0+3}}\right)\right\}$. \newline
It is nonhyperbolic with a three dimensional stable manifold provided
\newline
$\omega _0>-\frac{3}{2}, 1<l\leq \frac{1}{16} \left(6 \omega _0+25\right)$.

\item[$P_2$:] $(x,y,z,\lambda)=\left(-\sqrt{\frac{2}{3}},0,\sqrt{-\frac{2
\omega _0}{3}-1},0\right)$.\newline
There exists for $\omega _0<-\frac{3}{2}$. \newline
The eigenvalues are $\{-1,1,6,2-l\}$. \newline
It is always a saddle with a three dimensional unstable manifold if $l<2$.

\item[$P_3$:] $(x,y,z,\lambda)=\left(-\frac{\sqrt{6}}{l+1},-\frac{\sqrt{2-l}%
}{l+1},-\frac{\sqrt{(l-3) l-6 \omega _0-7}}{l+1},0\right)$. There exists for

\begin{enumerate}
\item $l<-1, \omega _0<-\frac{3}{2}$, or

\item $l<-1,  -\frac{3}{2}<\omega _0\leq \frac{1}{6} \left(l^2-3 l-7\right)$%
.
\end{enumerate}

The eigenvalues are \newline
$\Big\{-\frac{3}{l+1},6,\frac{3}{2 (l+1)}-\frac{\sqrt{6 (17-8 l)
\omega _0+8 l ((l-5) l-1)+121}}{2 (l+1) \sqrt{\frac{2 \omega _0}{3}+1}}$,\\
$\frac{3}{2 (l+1)}+\frac{\sqrt{6 (17-8 l) \omega _0+8 l ((l-5) l-1)+121}}{2
(l+1) \sqrt{\frac{2 \omega _0}{3}+1}}\Big\}$. \newline
It is a saddle.

\item[$P_4$:] $(x,y,z,\lambda)=\left(-\frac{\sqrt{6}}{l+1},\frac{\sqrt{2-l}}{%
l+1},\frac{\sqrt{(l-3) l-6 \omega _0-7}}{l+1},0\right)$. There exists for

\begin{enumerate}
\item $-1<l\leq 1, \omega _0<-\frac{3}{2}$ or

\item $-1<l\leq 1, -\frac{3}{2}<\omega _0\leq \frac{1}{6} \left(l^2-3
l-7\right)$,

\item $1<l<2, \omega _0\leq \frac{1}{6} \left(l^2-3 l-7\right)$ or

\item $l=2, \omega _0<-\frac{3}{2}$.
\end{enumerate}

The eigenvalues are\newline
$\Big\{-\frac{3}{l+1},6,\frac{3}{2 (l+1)}-\frac{\sqrt{6 (17-8 l)
\omega _0+8 l ((l-5) l-1)+121}}{2 (l+1) \sqrt{\frac{2 \omega _0}{3}+1}}$,\\
$\frac{3}{2 (l+1)}+\frac{\sqrt{6 (17-8 l) \omega _0+8 l ((l-5) l-1)+121}}{2
(l+1) \sqrt{\frac{2 \omega _0}{3}+1}}\Big\}$.  \newline
It is a saddle.

\item[$P_5$:] $(x,y,z,\lambda)=\left(\frac{2 \sqrt{\frac{2}{3}}}{2 \omega
_0+1},0,0,0\right)$. There is for

\begin{enumerate}
\item $\omega _0<-\frac{3}{2}$ or

\item $-\frac{5}{6}\leq \omega  _0<-\frac{1}{2}$ or

\item $\omega _0>-\frac{1}{2}$.
\end{enumerate}

The eigenvalues are $\left\{\frac{2}{2 \omega _0+1},\frac{2 (l+1)}{2 \omega
_0+1},-3,-\frac{2}{2 \omega _0+1}-3\right\}$. \newline
It is a sink for $l>-1, \omega _0<-\frac{3}{2}$. It is a saddle otherwise.

\item[$P_6$:] {\small{$(x,y,z,\lambda)=\left(-\frac{\sqrt{\frac{2}{3}} (l-1)}{l+2
\omega _0+2},-\frac{\sqrt{\left(2 \omega _0+3\right) \left(2 \omega _0-\frac{%
1}{3} (l-4) (l+2)\right)}}{l+2 \omega _0+2},0,0\right)$}}. \newline
There is for

\begin{enumerate}
\item $l<-2, \omega _0=\frac{1}{6} (l-4) (l+2)$ or

\item $l>1, \omega  _0=\frac{1}{6} (l-4) (l+2)$ or

\item $-2<l<1, \frac{1}{6} (l-4) (l+2)\leq  \omega _0<\frac{1}{2} (-l-2)$ or

\item $l\leq 1, \omega  _0<-\frac{3}{2}$ or

\item $l>1, \omega _0<\frac{1}{2} (-l-2)$.
\end{enumerate}

The eigenvalues are\\$\left\{-\frac{l-1}{l+2 \omega _0+2},\frac{2 (l-1) (l+1)%
}{l+2 \omega _0+2},\frac{l^2-3 l-6 \omega _0-7}{l+2 \omega _0+2},\frac{l^2-2
l-6 \omega _0-8}{l+2 \omega _0+2}\right\}$. \newline
It is a sink for $l<-1, \omega _0<-\frac{3}{2}$. It is a saddle otherwise.

\item[$P_7$:] {\small{$(x,y,z,\lambda)=\left(-\frac{\sqrt{\frac{2}{3}} (l-1)}{l+2
\omega _0+2},\frac{\sqrt{\left(2 \omega _0+3\right) \left(2 \omega _0-\frac{1%
}{3} (l-4) (l+2)\right)}}{l+2 \omega _0+2},0,0\right)$}}.\newline
There exists for

\begin{enumerate}
\item $l>1, \omega _0\geq \frac{1}{6}(l-4) (l+2)$ or

\item $l<-2,  \omega _0\geq \frac{1}{6} (l-4) (l+2)$ or

\item $l=-2, \omega  _0>0$ or

\item $l=1, \omega _0>-\frac{3}{2}$ or

\item $-2<l<1, \omega  _0>\frac{1}{2} (-l-2)$ or

\item $l>1, \frac{1}{2} (-l-2)<\omega  _0<-\frac{3}{2}$ or

\item $-2<l<1, \omega _0=\frac{1}{6} (l-4) (l+2)$.
\end{enumerate}

The eigenvalues are\\$\left\{-\frac{l-1}{l+2 \omega _0+2},\frac{2 (l-1) (l+1)%
}{l+2 \omega _0+2},\frac{l^2-3 l-6 \omega _0-7}{l+2 \omega _0+2},\frac{l^2-2
l-6 \omega _0-8}{l+2 \omega _0+2}\right\}$.  \newline
It is a saddle.

\item[$P_8$:] $(x,y,z,\lambda)=\left(\frac{\sqrt{3}-\sqrt{2 \omega _0+3}}{%
\sqrt{2} \omega _0},0,0,0\right)$. \newline
There is for $\omega _0>-\frac{3}{2}, \omega _0\neq 0$. \newline
The eigenvalues are \newline
$\Big\{-\frac{\sqrt{6 \omega _0+9}-3}{2 \omega _0},-\frac{\sqrt{6 \omega
_0+9}-3}{2 \omega _0},\frac{6 \omega _0-\sqrt{6 \omega _0+9}+3}{\omega _0}, \frac{6 \omega _0-(l+2) \left(\sqrt{6 \omega _0+9}-3\right)}{2 \omega _0}%
\Big\}$.  \newline
It is a saddle.

\item[$P_9$:] $(x,y,z,\lambda)=\left(\frac{\sqrt{3}+\sqrt{2 \omega _0+3}}{%
\sqrt{2} \omega _0},0,0,0\right)$. \newline
There exists for $\omega _0>-\frac{3}{2}, \omega _0\neq 0$.\newline
The eigenvalues are\newline
$\Big\{\frac{\sqrt{6 \omega _0+9}+3}{2 \omega _0},\frac{\sqrt{6 \omega _0+9}%
+3}{2 \omega _0},\frac{6 \omega _0+\sqrt{6 \omega _0+9}+3}{\omega _0}, \frac{%
(l+2) \left(\sqrt{6 \omega _0+9}+3\right)+6 \omega _0}{2 \omega _0}\Big\}$%
. \newline
It is a sink for

\begin{enumerate}
\item $-2<l\leq -1, \frac{1}{6} (l-4) (l+2)<\omega _0<0$, or

\item $l>-1, -\frac{5}{6}<\omega _0<0$.
\end{enumerate}

It is a source for

\begin{enumerate}
\item $l\leq -2, \omega _0>\frac{1}{6} (l-4) (l+2)$, or

\item $l>-2, \omega_0>0$. \newline
It is a saddle otherwise.
\end{enumerate}

\item[$P_{10}$:]  $(x,y,z,\lambda)=\left(0,0,\sqrt{\frac{2}{3}},2\right)$.
 It always exists. The eigenvalues are \newline
$\left\{-1,2,\frac{1}{2} \left(-\sqrt{1-\frac{48}{2 \omega _0+3}}-1\right),%
\frac{1}{2}  \left(\sqrt{1-\frac{48}{2 \omega _0+3}}-1\right)\right\}$.%
\newline
It is a saddle with a three dimensional stable manifold provided $\omega
_0\geq \frac{45}{2}$.
\end{enumerate}

\subsubsection{Center manifold of $P_1$.}
\label{Center_manifolds_P1_a}

From the previous linear analysis, we have found that the equilibrium point $P_1$ is nonhyperbolic with a three dimensional stable manifold provided
$\omega _0>-\frac{3}{2}, 1<l\leq \frac{1}{16} \left(6 \omega _0+25\right)$.
In this subsection, we use the Center Manifold Theorem to show that the solution corresponding to $P_1$ is indeed locally asymptotically stable under the above conditions.

Introducing the new variables:
\begin{widetext}
\begin{subequations}
\begin{align}
&u=\lambda,\\
& v_1=z-\frac{\lambda  (l-1)}{\sqrt{6}
   (l+1)},\\
&  v_2= \frac{-\frac{x
   \left(4 (l-4) \omega _0+6 l\right)}{\sqrt{l+1}}+\left(y-\frac{\sqrt{2}}{\sqrt{l+1}}\right) \left(\sqrt{\left(2 \omega
   _0+3\right) \left(-16 l+6 \omega _0+25\right)}-2 \sqrt{3} \omega _0+5 \sqrt{3}\right)}{2 \sqrt{\left(2 \omega _0+3\right) \left(-16
   l+6 \omega _0+25\right)}},\\
 & v_3= \frac{\frac{x \left(4 (l-4) \omega _0+6
   l\right)}{\sqrt{l+1}}-\left(y-\frac{\sqrt{2}}{\sqrt{l+1}}\right) \left(-\sqrt{\left(2 \omega
   _0+3\right) \left(-16 l+6 \omega _0+25\right)}-2 \sqrt{3} \omega _0+5 \sqrt{3}\right)}{2
   \sqrt{\left(2 \omega _0+3\right) \left(-16 l+6 \omega _0+25\right)}},
\end{align}
\end{subequations}
\end{widetext}

which are real,  the point $P_1$ is  shifted to the origin  and the  linear part of the vector field is transformed to its real Jordan canonical form.
Therefore, the evolution equations becomes
\begin{equation}
\label{JBDcenter2}
\left(\begin{array}{c} u'\\v_1'\\v_2'\\v_3'
\end{array}\right)=\left(\begin{array}{ccccc} 0& 0 &0 &0  \\
0& -3 &0 &0
\\  0& 0 & \lambda_2 &0
\\ 0& 0 & 0 &  \lambda_3 \\
\end{array}\right)\left(\begin{array}{c}u\\v_1\\v_2\\v_3
\end{array}\right)+\left(\begin{array}{c}f(u,\mathbf{v})\\g_1(u,\mathbf{v})\\g_2(u,\mathbf
{v})
\\g_3(u,\mathbf{v})\end{array}\right)
\end{equation}
 where
\begin{subequations}
\begin{align}
& \lambda_2=  -\frac{ \left(\sqrt{3} \sqrt{\left(2 \omega _0+3\right) \left(-16 l+6
   \omega _0+25\right)}+6 \omega _0+9\right)}{4 \omega _0+6},\\
& \lambda_3=  -\frac{\left(6 \omega _0+9 - \sqrt{3}
   \sqrt{\left(2 \omega _0+3\right) \left(-16 l+6 \omega _0+25\right)}\right)}{4
   \omega _0+6},
	\end{align}
	\begin{widetext}
	\begin{align}
&   f(u,\mathbf{v})=\frac{\sqrt{l+1} u  \left(\sqrt{3} \sqrt{\left(2 \omega _0+3\right)
   \left(-16 l+6 \omega _0+25\right)} (v_3-v_2)-6 \omega _0 (v_2+v_3)+15
   (v_2+v_3)\right)}{\sqrt{2} \left(4 (l-4) \omega _0+6 l\right)}
   \end{align}
	\end{widetext}
	\end{subequations}
\\ and the
$g_1(u,\mathbf{v}), g_2(u,\mathbf{v}), g_3(u,\mathbf{v})$ are more complicated expressions. That is, 
\begin{align}
u'  =Cu+f\left(  u,\mathbf{v}\right), \quad
\mathbf{v}'  =P\mathbf{v}+\mathbf{g}\left(  u,\mathbf{v}\right)
, \label{JBDcenter3}
\end{align}
where $\left(  u,\mathbf{v}\right)
\in\mathbb{R}\times\mathbb{R}^{3},$ $C\equiv 0 \in \mathbb{R}^{1\times 1}$, $P \in \mathbb{R}^{3\times 3}$ has negative eigenvalues, and  $f(0)=0, \quad \mathbf{g}(\mathbf{0})=\mathbf{0}$, $f'(0)=0, \quad D\mathbf{g}(\mathbf{0})=\mathbf{0}$. Using the center manifold theorem we have that there exists a 1-dimensional
invariant local center manifold $W^{c}\left( \mathbf{0}\right) $
of \eqref{JBDcenter3} tangent to the center subspace $\mathbf{v}=\mathbf{0}$ such that
\begin{align}
&W^{c}\left(  \mathbf{0}\right)  =\Big\{  \left(
u,\mathbf{v}\right)
\in\mathbb{R}\times\mathbb{R}^{3}:\mathbf{v}=\mathbf{h}\left(
u\right), \nonumber\\
\;\;\;\;\;\;\;\;\;\;\;\;\;\;\;\;\;\;\;\;\;\;\;\; & \mathbf{h}\left(  0\right) =\mathbf{0},\;D\mathbf{h}\left(
0\right)  =\mathbf{0},\;\left\vert u\right\vert <\delta\Big\},
\end{align} 
where $\delta$ is small enough. 
The dynamics on the center manifold is governed by the equation
\begin{equation}
u'=f\left( u,\mathbf{h}\left(  u\right)  \right)  . \label{JBDrest}
\end{equation}
According to the center manifold theorem, if the origin of
\eqref{JBDrest} is stable, asymptotically stable or unstable,  then
the origin of \eqref{JBDcenter3} is also stable, asymptotically
stable or unstable. 

Completing the analysis is required the computation of
$\mathbf{h}\left( u\right).$ By substituting $\mathbf{v}=\mathbf{h}\left(  u\right)$ in 
\eqref{JBDcenter3} and using the chain rule we can deduce the system of differential equations
\begin{equation}
D\mathbf{h}\left(  u\right)  \left[  f\left(  u,\mathbf{h}\left(
u\right) \right)  \right]  -P\mathbf{h}\left(  u\right)
-\mathbf{g}\left( u,\mathbf{h}\left(  u\right)  \right)  =0.
\label{JBDh}
\end{equation}
The system \eqref{JBDh}
can be solved approximately by taking Taylor series of 
$\mathbf{h}\left(u\right)$ centered in $u=0.$ Since
$\mathbf{h}\left(  0\right)  =\mathbf{0\ }$ and $D\mathbf{h}\left(
0\right)  =\mathbf{0},$ one substitutes
\[
\mathbf{h}\left(  x\right)  :=\left[
\begin{array}
[c]{c}%
h_{1}\left(  u\right) \\
h_{2}\left(  u\right) \\
h_{3}\left(  u\right)
\end{array}
\right]  =\left[
\begin{array}
[c]{c}%
a_{1}u^{2}+\mathcal{O}\left(  u^{3}\right) \\
a_{2}u^{2}+\mathcal{O}\left(  u^{3}\right) \\
a_{3}u^{2}+\mathcal{O}\left(  u^{3}\right)
\end{array}
\right]
\]
into (\ref{JBDh}). Collecting the coefficients of the same powers of $u$ in the left hand side of (\ref{JBDh}), and setting them to zero, we  obtain the non-trivial
coefficients
\begin{widetext}
\begin{align*}
&a_2= -\frac{\sqrt{2} (l-1)^2 \left(\omega _0 \sqrt{\left(2 \omega
   _0+3\right) \left(-16 l+6 \omega _0+25\right)}+\sqrt{3} l \left(2 \omega _0+3\right)-\sqrt{3}
   \left(2 \omega _0+3\right) \omega _0\right)}{(l+1)^{5/2} \left(2 \omega _0+3\right) \left(-3
   \sqrt{\left(2 \omega _0+3\right) \left(-16 l+6 \omega _0+25\right)}+16 \sqrt{3} l-\sqrt{3}
   \left(6 \omega _0+25\right)\right)},\\
&   a_3= -\frac{\sqrt{2} (l-1)^2 \left(-\omega _0
   \sqrt{\left(2 \omega _0+3\right) \left(-16 l+6 \omega _0+25\right)}+\sqrt{3} l \left(2 \omega
   _0+3\right)-\sqrt{3} \left(2 \omega _0+3\right) \omega _0\right)}{(l+1)^{5/2} \left(2 \omega
   _0+3\right) \left(3 \sqrt{\left(2 \omega _0+3\right) \left(-16 l+6 \omega _0+25\right)}+16
   \sqrt{3} l-\sqrt{3} \left(6 \omega _0+25\right)\right)}.
\end{align*}
\end{widetext}
\begin{figure*}[t]
\centering
{\includegraphics[width=0.6\textwidth]{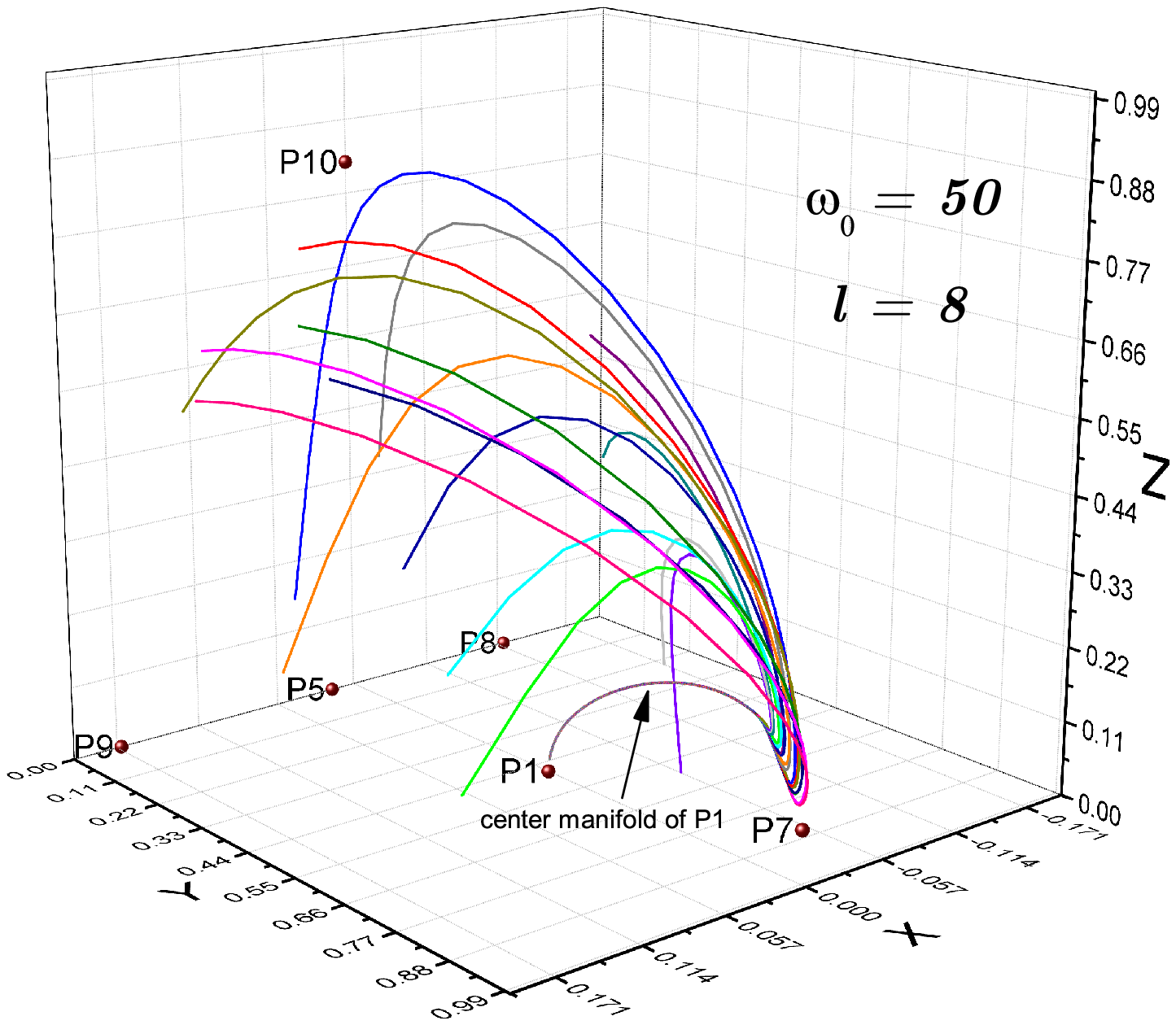}}
\caption{ \label{FIG1} (Color online) \emph {Case: $W(\protect\psi)=W_0 e^{k \protect\psi}$ and $V(\protect%
\varphi)=V_0 e^{l \protect\varphi}$. Evolution of some orbits of the dynamical system \eqref{syst1exp_exp} projected on space $(x,y,z)$ for $\omega_0=50, l=8.$ The initial conditions are chosen randomly to show that, irrespectively of the initial conditions, the orbits are attracted by the center manifold of the equilibrium point $P_1$.}}
\bigskip
\centering
{\includegraphics[width=0.6\textwidth]{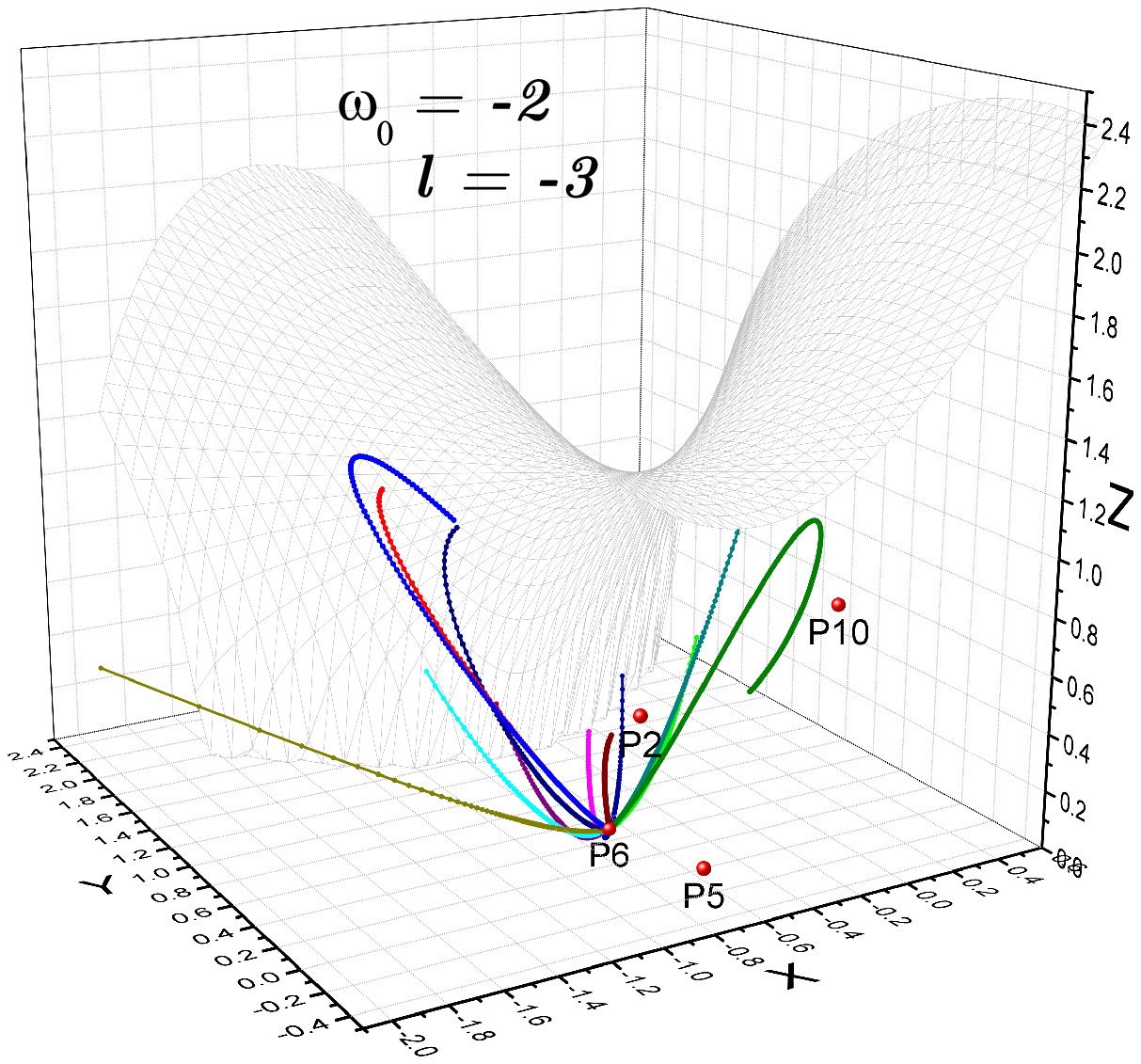}}
\caption{ \label{FIG2} (Color online) \emph {Case: $W(\protect\psi)=W_0 e^{k \protect\psi}$ and $V(\protect%
\varphi)=V_0 e^{l \protect\varphi}$. Evolution of some orbits of the dynamical system \eqref{syst1exp_exp} projected on space $(x,y,z)$ for $\omega_0=-2, l=-3.$ The initial conditions are chosen randomly to show that, irrespectively of the initial conditions, the orbits are attracted by the the equilibrium point $P_6$.}}
\end{figure*}

Therefore, the local center manifold of the origin can be expressed
\begin{widetext}
\begin{align*}
&\left\{(u, v_1, v_2, v_3)\in \mathbb{R}^{4}: v_1=0, \right. \nonumber \\
& \left. v_2=-\frac{\sqrt{2} (l-1)^2 \left(-\sqrt{3} \left(2 \omega _0+3\right)
   \omega _0+\sqrt{\left(2 \omega _0+3\right) \left(-16 l+6 \omega _0+25\right)} \omega
   _0+\sqrt{3} l \left(2 \omega _0+3\right)\right) u^2}{(l+1)^{5/2} \left(2 \omega _0+3\right)
   \left(16 \sqrt{3} l-\sqrt{3} \left(6 \omega _0+25\right)-3 \sqrt{\left(2 \omega _0+3\right)
   \left(-16 l+6 \omega _0+25\right)}\right)}, \right. \nonumber \\
& \left. v_3= -\frac{\sqrt{2} (l-1)^2
   \left(-\sqrt{3} \left(2 \omega _0+3\right) \omega _0-\sqrt{\left(2 \omega _0+3\right)
   \left(-16 l+6 \omega _0+25\right)} \omega _0+\sqrt{3} l \left(2 \omega _0+3\right)\right)
   u^2}{(l+1)^{5/2} \left(2 \omega _0+3\right) \left(16 \sqrt{3} l-\sqrt{3} \left(6 \omega
   _0+25\right)+3 \sqrt{\left(2 \omega _0+3\right) \left(-16 l+6 \omega_0+25\right)}\right)}\right\}.
\end{align*}
\end{widetext}

The dynamics on the center manifold are given by a gradient like equation $
u'=-\nabla \Pi(u)$, where $\Pi(u)=\frac{(l-1) u^4}{8 (l+1)^2}$, 
for which the origin is a degenerate local minimum whenever $l>1$ (recall the existence conditions for $P_1$ are $2 \omega _0+3\neq 0, l\geq 1$). This implies that the center manifold of $P_1$ is stable when $1<l\leq \frac{1}{16} \left(6 \omega _0+25\right)$. For $l> \frac{1}{16} \left(6 \omega _0+25\right)$ the unstable manifold is not empty. Neglecting the order terms $\mathcal{O}(\lambda^3)$ the center can be given in the original variables by the graph:
\begin{subequations}
\begin{align}
\label{Center_coordinates}
 x= -\frac{(l-1) \lambda ^2}{\sqrt{6} (l+1)^2},  \quad 
 y=\frac{\sqrt{2}}{\sqrt{l+1}}-\frac{\left((l-1) \left(l-2 \omega _0\right)\right) \lambda ^2}{4\left(\sqrt{2} (l+1)^{5/2}\right)}, 
\quad  z= \frac{(l-1) \lambda}{\sqrt{6} (l+1)}.
\end{align}
\end{subequations}
In the figure \ref{FIG1} we present some orbits of the dynamical system \eqref{syst1exp_exp} projected on the space $(x,y,z)$  for $W(\protect\psi)=W_0 e^{k \protect\psi}$ and $V(\protect%
\varphi)=V_0 e^{l \protect\varphi}$ with $\omega_0=50$ and $l=8$. The initial conditions were chosen randomly to show that, irrespectively of the initial conditions, the orbits are attracted by the center manifold of the equilibrium point $P_1$. Latter on, in Section \ref{Sect:6.1} it will be shown that the cosmological solutions  represented by these orbits, tends to the solution associated to $P_1$. Furthermore, as it is shown in Figure \ref{FIG1b}, the cosmological parameters behave in accordance with the current cosmological paradigm. This feature makes the model very interesting from the cosmological point of view.


In the Fig. \ref{FIG2} are displayed  some orbits of the dynamical system \eqref{syst1exp_exp} projected on space $(x,y,z)$ for $\omega_0=-2, l=-3.$ The initial conditions are chosen randomly to show that, irrespectively of the initial conditions, the orbits are attracted by the  equilibrium point $P_6$.
In this example, the phase space is the interior of a hyperboloid that corresponds to the boundary of the phase space and it is represented by a gray mesh. The late-time attractor is a phantom dominated solution.

As we have commented before, the model studied in \cite{Cid:2015pja} can be considered as a special case of the model  $W(\protect\psi)=W_0 e^{k \protect\psi}$ and $V(\protect%
\varphi)=V_0 e^{l \protect\varphi}$ with\\$k=-\lambda_W, l=1-\lambda_U \sqrt{\omega_0+\frac{3}{2}}, \omega_0>-\frac{3}{2}$. 
In this section, we have investigated the stability of the equilibrium solutions in the dilatonic frame. In the reference \cite{Cid:2015pja} it was studied the stability of the equilibrium points in both the Jordan and the Einstein frames, so our results complement those found in  \cite{Cid:2015pja}. 
In particular, we notice that the equilibrium point (in the Jordan frame), named $J_4$ in \cite{Cid:2015pja}  corresponds to $P_1$ investigated in this section with the identification $\lambda_U=(1-l)\gamma, \gamma^{-1}=\sqrt{\omega_0+\frac{3}{2}}$, due to it satisfies 
\begin{align}
& J_4: \left(e^{-\frac{\varphi}{2}}\frac{\dot \psi}{\sqrt{6} H }, \frac{\dot \varphi}{H}, \frac{\sqrt{V(\varphi)}}{\sqrt{3} H}\right) =\left(0,0,\frac{\sqrt{2 \gamma}}{\sqrt{2 \gamma -\lambda_U}}\right)=\left(0,0,\frac{\sqrt{2}}{\sqrt{1+l}}\right).
\end{align}
The stability conditions deduced in  \cite{Cid:2015pja} the Jordan's frame formulation and also in the Einstein's frame formulation are $\lambda_U<0, \gamma>0$. That is, $\omega_0>-\frac{3}{2}, l>1$. The stability in the dilatonic frame formulation is $1<l\leq \frac{1}{16} \left(6 \omega _0+25\right)$. They are the equivalent conditions with the identifications $\lambda_U=(1-l)\gamma, \gamma^{-1}=\sqrt{\omega_0+\frac{3}{2}}$. The equilibrium points $J_4$ and $E_4$ (the representations of $P_1$ in the Jordan's frame and in the Einstein's frame, respectively) correspond to an intermediate accelerated solution instead of a de Sitter solution (see derivation in \cite{Cid:2015pja}). That is, an attractor in the Jordan frame corresponds to the solution of the form $a(t) \simeq e^{\alpha_1 t^{p_1}}$, as $t\rightarrow \infty$ where $\alpha_1 > 0$ and $0 < p_1 < 1$ for a wide range of parameters. Furthermore, when we work in the Einstein frame we get that the attractor is also the solution of the form $\bar{a}(\bar{t}) \simeq e^{\alpha_2 \bar{t}^{p_2}}$, as $\bar{t}\rightarrow \infty$ where $\alpha_2 > 0$ and $0<p_2<1$, for the same conditions on the parameter space as in the Jordan frame. An equivalent result  can be deduced straightforwardly for the dilatonic frame.  We proceed as follows.
According to the center manifold calculation, we have from \eqref{Center_coordinates}, the definition $\lambda:=k e^{-\varphi/2}$, and the definition \eqref{coordinates-x-y-z}
that (as $\varphi\rightarrow \infty$):
\begin{subequations}
\begin{align}
 \frac{d \varphi}{d \ln a}=-\frac{(l-1) k^2 e^{-\varphi}}{ (l+1)^2}, \  \frac{d \psi}{d \ln a}=\frac{(l-1) k
   }{(l+1)}, \	\frac{d t}{d \ln a}= \frac{8 (l+1)^2 e^{\varphi }-k^2 (l-1) (l-2 \omega_{0})}{4 \sqrt{\frac{2}{3}} (l+1)^{5/2} \sqrt{V_{0}} e^{\frac{(2-l)
   \varphi }{2}}}.
\end{align}
\end{subequations}
With general solution 
\begin{subequations}
\begin{align}
 \varphi (a)= \ln \left|c_1-\frac{k^2 (l-1) \ln(a)
   }{(l+1)^2}\right|, \quad \psi (a)= c_3+\frac{k (l-1) \ln(a)
   }{l+1}
   ,\end{align}
\begin{align}
& t(a)= \frac{\sqrt{\frac{3}{2}} \left|c_1-\frac{k^2 (l-1)
   \ln(a)}{(l+1)^2}\right|{}^{l/2} \left(k^2 (l-1) (l (l+8 \ln(a)+2)-2 (l+2)\omega_{0})-8 c_1 l (l+1)^2\right)}{2 k^2 (l-1) l \sqrt{l+1} (l+2)
   \sqrt{V_{0}}}+c_2,
\end{align}
\end{subequations}
where $c_1, c_2, c_3$ are integration constants.
For large $a$, the leading terms are 
\begin{align}
& t(a)\simeq\frac{2 \sqrt{6} \ln(a) 
   \left|c_1-\frac{k^2 (l-1) \ln(a)}{(l+1)^2}\right|{}^{l/2}}{\sqrt{l+1} (l+2) \sqrt{V_{0}}} \nonumber \\
	& \implies 
\ln a \simeq t^{\frac{2}{2+l}}\implies a \simeq e^{A t^p}, \quad p:=\frac{2}{2+l}, \quad \frac{32}{57+6 \omega_0}<p<\frac{2}{3}, \;\text{as}\; t\rightarrow \infty.
\end{align}
with the identifications  $\lambda_U=(1-l)\gamma, \gamma^{-1}=\sqrt{\omega_0+\frac{3}{2}}$ we obtain the same exponent $p=p_1=p_2=\frac{2 \gamma}{3\gamma-\lambda_U}$. Since $p<\frac{2}{3}$, $P_1$ is not a de Sitter solution (that requires $p=1$).

\subsection{Case: $W(\protect\psi)=W_0 \protect\psi^{k}$ and $V(\protect%
\varphi)=V_0 e^{l \protect\varphi}$}

\label{sect:4.2} In this example we have $s=-l=\text{const.},\lambda =-\frac{%
k e^{\varphi /2}}{\psi },\Gamma_\varphi =1,\Gamma_\psi =\frac{k-1}{k}$. Thus
$h(s,\lambda)=\frac{k-1}{k}$. Assuming $k\neq 0$ and introducing $x_1=\frac{%
2 \lambda z}{k}+x$ the system \eqref{syst1} becomes

\begin{subequations}
\label{syst1:power-exp}
\begin{align}
&x_1^{\prime }=\Big[-\frac{3 (l+1) y^2}{2
\omega _0+3}+\frac{6 \lambda ^2}{k}+z \left(\frac{4 \sqrt{6} \omega _0
\lambda ^3}{k^2}+\frac{12 \sqrt{6} \left(2 \omega _0+1\right) \lambda }{k
\left(2 \omega _0+3\right)}\right) \notag \\
&  +z^2 \left(\frac{12 \omega _0 \left(2 \omega _0+1\right) \lambda ^2}{k^2
\left(2 \omega _0+3\right)}+\frac{6 \omega _0}{2 \omega _0+3}\right) +\frac{%
18}{2 \omega _0+3}-3\Big] x_1  \notag \\
& +\left[-\frac{\sqrt{6} \omega _0 \lambda ^2}{k}-\frac{12 z \omega _0
\left(2 \omega _0+1\right) \lambda }{k \left(2 \omega _0+3\right)}-\frac{%
\sqrt{6} \left(8 \omega _0+3\right)}{2 \omega _0+3}\right] x_1^2 \notag \\
& + \left[3 \omega _0+\frac{9}{2 \omega _0+3}-3\right] x_1^3 +y^2 \left[-%
\frac{\sqrt{6} \lambda ^2}{k}-\frac{\sqrt{6} (l+1)}{2 \omega _0+3}\right]   \notag \\
& +z^2 \left[-\frac{4 \sqrt{6} \omega _0 \lambda ^4}{k^3}-\frac{\sqrt{6}
\left(3 (k+2)+2 (k+6) \omega _0\right) \lambda ^2}{k^2 \left(2 \omega
_0+3\right)}-\frac{3 \sqrt{6}}{2 \omega _0+3}\right] \notag \\
&  +z \left[-\frac{12 \lambda ^3}{k^2}-\frac{24 \lambda }{2 \omega _0 k+3 k}%
\right]  +\frac{2 \sqrt{6}}{2 \omega _0+3}+%
\frac{\sqrt{6} \lambda ^2}{k} ,\end{align}
\begin{align}
& y^{\prime }= \left[\frac{6}{2 \omega _0+3}\right] y+\left[\frac{\sqrt{6}
\left(3 l+2 (l-4) \omega _0\right)}{4 \omega _0+6}-\frac{12 z \lambda \omega
_0 \left(2 \omega _0+1\right)}{k \left(2 \omega _0+3\right)}\right] x_1 y
\notag \\
& +\left[\left(\frac{12 \omega _0 \left(2 \omega _0+1\right) \lambda ^2}{k^2
\left(2 \omega _0+3\right)}+\frac{6 \omega _0}{2 \omega _0+3}\right) z+\frac{%
\sqrt{6} \lambda \left(-3 l-2 (l-4) \omega _0\right)}{k \left(2 \omega
_0+3\right)}\right] z y  \notag \\
& -\frac{3 (l+1) y^3}{2 \omega _0+3}+ \left[3 \omega _0+\frac{9}{2 \omega
_0+3}-3\right]x_1^2 y,
\end{align}
\begin{align}
& z^{\prime }=\sqrt{\frac{3}{2}} \lambda+\left[\frac{12 \omega _0 \left(2
\omega _0+1\right) \lambda ^2}{k^2 \left(2 \omega _0+3\right)}+\frac{6
\omega _0}{2 \omega _0+3}\right] z^3  \notag \\
&  +\left[\frac{\sqrt{\frac{3}{2}} \lambda \left(-3 k-2 (k-10) \omega
_0+6\right)}{k \left(2 \omega _0+3\right)}-\frac{2 \sqrt{6} \lambda ^3
\omega _0}{k^2}\right] z^2  \notag \\
& +\left[-\frac{6 \lambda ^2}{k}+\frac{6}{2 \omega _0+3}-3\right] z+y^2 %
\left[-\frac{3 (l+1) z}{2 \omega _0+3}-\sqrt{\frac{3}{2}} \lambda \right]
\notag \\
& +x_1^2 \left[z \left(3 \omega _0+\frac{9}{2 \omega _0+3}-3\right)-\sqrt{%
\frac{3}{2}} \lambda \omega _0\right]  \notag \\
& +x_1 \left[-\frac{12 \lambda \omega _0 \left(2 \omega _0+1\right) z^2}{k
\left(2 \omega _0+3\right)}+\left(\frac{2 \sqrt{6} \lambda ^2 \omega _0}{k}-%
\frac{\sqrt{6} \left(10 \omega _0+3\right)}{4 \omega _0+6}\right) z+3
\lambda \right],\\
& \lambda^{\prime }=\sqrt{\frac{3}{2}} x_1 \lambda.
\end{align}
\end{subequations}
The system is a form-invariant under the change $(y,
z,\lambda)\rightarrow (-y,-z, -\lambda)$. Therefore, without losing generality we
can investigate just the sector $y\geq 0, z\geq 0, \lambda\geq 0$.
Henceforth, we will focus on the stability properties of the system %
\eqref{syst1:power-exp} for the state vector $(x,y,z, \lambda)$ defined in
the phase space
\begin{align}
&\Big\{(x_1,y,z, \lambda)\in\mathbb{R}^4: y\geq 0, z\geq 0, \lambda \geq 0, 2 \omega
_0+3\neq 0,  k \left(k \left(-\sqrt{6} x_1+y^2+z^2-1\right)+2 \sqrt{6} \lambda
z\right)  \notag \\
& +\omega _0 (k x_1-2 \lambda z)^2\leq 0\Big\}.
\end{align}

We will focus on the study of the particular choice of potentials \eqref{mn.06}%
:
\begin{equation}
V\left( \varphi \right) =V_{0}e^{\left( \beta -1\right) \varphi }~\
,~W\left( \psi \right) =W_{0}\psi ^{2\beta },
\end{equation}
that lead to Noether pointlike symmetries, corresponding to the choice $%
l=\beta-1$, and $k=2\beta$. The equilibrium points are the following

\begin{enumerate}
\item[$P_1$] $:=(x_1,y,z,\lambda)=\left(0,\frac{\sqrt{2}}{\sqrt{\beta }}%
,0,0\right)$. \newline
The eigenvalues are \newline
$\Big\{0,-3,-\frac{\sqrt{3} \sqrt{\left(2 \omega _0+3\right) \left(-16
\beta +6 \omega _0+41\right)}}{4 \omega _0+6}-\frac{3}{2},\frac{1}{2} \left(%
\frac{\sqrt{3} \sqrt{\left(2 \omega _0+3\right) \left(-16 \beta +6 \omega
_0+41\right)}}{2 \omega _0+3}-3\right)\Big\}$. \newline
The stable manifold is three dimensional for $\beta >2, \omega _0\geq \frac{1%
}{6} (16 \beta -41)$.

\item[$P_2$] $:= (x_1,y,z,\lambda)=\left(-\sqrt{\frac{2}{3}},0,\sqrt{-\frac{2
\omega _0}{3}-1},0\right)$. 
The eigenvalues are \newline
$\{-1,1,6,3-\beta \}$.  \newline
It is a saddle point with a three dimensional unstable manifold for $\beta<3$%
.

\item[$P_3$] $:= (x_1,y,z,\lambda)=\left(-\frac{\sqrt{6}}{\beta },-\frac{\sqrt{%
3-\beta }}{\beta },-\frac{\sqrt{(\beta -5) \beta -6 \omega _0-3}}{\beta }%
,0\right)$. \newline
The eigenvalues are 
$\Big\{-\frac{3}{\beta },6,\frac{3}{2 \beta }-\frac{\sqrt{6 (25-8 \beta )
\omega _0+8 (\beta -6) (\beta -2) \beta +81}}{2 \beta \sqrt{\frac{2 \omega _0%
}{3}+1}}$,\\
$\frac{3}{2 \beta }+\frac{\sqrt{6 (25-8 \beta ) \omega _0+8 (\beta
-6) (\beta -2) \beta +81}}{2 \beta \sqrt{\frac{2 \omega _0}{3}+1}}\Big\}$.
\newline
It is a saddle.

\item[$P_4$] $:= (x_1,y,z,\lambda)=\left(-\frac{\sqrt{6}}{\beta },\frac{\sqrt{%
3-\beta }}{\beta },\frac{\sqrt{(\beta -5) \beta -6 \omega _0-3}}{\beta }%
,0\right)$.\newline
The eigenvalues are 
$\Big\{-\frac{3}{\beta },6,\frac{3}{2 \beta }-\frac{\sqrt{6 (25-8 \beta )
\omega _0+8 (\beta -6) (\beta -2) \beta +81}}{2 \beta \sqrt{\frac{2 \omega _0%
}{3}+1}}$,\\
$\frac{3}{2 \beta }+\frac{\sqrt{6 (25-8 \beta ) \omega _0+8 (\beta
-6) (\beta -2) \beta +81}}{2 \beta \sqrt{\frac{2 \omega _0}{3}+1}}\Big\}$.
\newline
It is a saddle.

\item[$P_5$] $:= (x_1,y,z,\lambda)=\left(\frac{2 \sqrt{\frac{2}{3}}}{2 \omega
_0+1},0,0,0\right)$. \newline
The eigenvalues are $\left\{\frac{2}{2 \omega _0+1},\frac{2 \beta }{2 \omega
_0+1},-3,-\frac{6 \omega _0+5}{2 \omega _0+1}\right\}$.  \newline
It is a sink for $\beta >0, \omega _0<-\frac{3}{2}$.

\item[$P_6$] $:= (x_1,y,z,\lambda)$\\
$=\left(-\frac{\sqrt{\frac{2}{3}} (\beta -2)}{%
\beta +2 \omega _0+1},-\frac{\sqrt{\left(2 \omega _0+3\right) \left(2 \omega
_0-\frac{1}{3} (\beta -5) (\beta +1)\right)}}{\beta +2 \omega _0+1}%
,0,0\right)$.  \newline
The eigenvalues are \newline
$\left\{-\frac{\beta -2}{\beta +2 \omega _0+1},\frac{2 (\beta -2) \beta }{%
\beta +2 \omega _0+1},\frac{\beta ^2-5 \beta -6 \omega _0-3}{\beta +2 \omega
_0+1},\frac{\beta ^2-4 \beta -6 \omega _0-5}{\beta +2 \omega _0+1}\right\}$.
\newline
It is a sink for $\beta <0, \omega _0<-\frac{3}{2}$.  It is a saddle
otherwise.

\item[$P_7$] $:=(x_1,y,z,\lambda)$\\
$=\left(-\frac{\sqrt{\frac{2}{3}} (\beta -2)}{%
\beta +2 \omega _0+1},\frac{\sqrt{\left(2 \omega _0+3\right) \left(2 \omega
_0-\frac{1}{3} (\beta -5) (\beta +1)\right)}}{\beta +2 \omega _0+1}%
,0,0\right)$.  \newline
The eigenvalues are \newline
$\left\{-\frac{\beta -2}{\beta +2 \omega _0+1},\frac{2 (\beta -2) \beta }{%
\beta +2 \omega _0+1},\frac{\beta ^2-5 \beta -6 \omega _0-3}{\beta +2 \omega
_0+1},\frac{\beta ^2-4 \beta -6 \omega _0-5}{\beta +2 \omega _0+1}\right\}$.
\newline
It is a saddle.

\item[$P_8$] $:= (x_1,y,z,\lambda)=\left(\frac{\sqrt{3}-\sqrt{2 \omega _0+3}}{%
\sqrt{2} \omega _0},0,0,0\right)$.\newline
The eigenvalues are \newline
$\Big\{-\frac{\sqrt{6 \omega _0+9}-3}{2 \omega _0},-\frac{\sqrt{6 \omega
_0+9}-3}{2 \omega _0},\frac{6 \omega _0-\sqrt{6 \omega _0+9}+3}{\omega _0},\frac{6 \omega _0-(\beta +1) \left(\sqrt{6 \omega _0+9}-3\right)}{2 \omega _0%
}\Big\}$.\newline
It is a saddle.

\item[$P_9$] $:=(x_1,y,z,\lambda)=\left(\frac{\sqrt{3}+\sqrt{2 \omega _0+3}}{%
\sqrt{2} \omega _0},0,0,0\right)$.\newline
The eigenvalues are 
$\Big\{\frac{\sqrt{6 \omega _0+9}+3}{2 \omega _0},\frac{\sqrt{6 \omega _0+9}%
+3}{2 \omega _0},\frac{6 \omega _0+\sqrt{6 \omega _0+9}+3}{\omega _0}$,\\
$\frac{%
(\beta +1) \left(\sqrt{6 \omega _0+9}+3\right)+6 \omega _0}{2 \omega _0}%
\Big\}$.\newline
It is a sink for

\begin{enumerate}
\item $-1<\beta \leq 0, \frac{1}{6} (\beta -5) (\beta +1)<\omega _0<0$, or

\item $\beta >0, -\frac{5}{6}<\omega _0<0$.
\end{enumerate}

It is a source for

\begin{enumerate}
\item $\beta \leq -1, \omega _0>\frac{1}{6} (\beta -5) (\beta +1)$, or

\item $\beta  >-1, \omega _0>0$.
\end{enumerate}

It is a saddle otherwise.
\begin{widetext}
\item[$P_{11}(\lambda)$] $:=(x_1,y,z,\lambda)=\Big(0, -\frac{\sqrt{2 \lambda ^2 \omega _0 \left((14-(\beta -4) \beta )
\lambda ^2+6 \beta (\beta +2)+6 \lambda ^2 \omega _0\right)-3 \left((\beta
-5) \lambda ^2-4 \beta \right) \left(2 \beta ^2+(\beta +1) \lambda ^2\right)}%
}{\sqrt{3} \left(2 \beta ^2+(\beta +1) \lambda ^2+2 \lambda ^2 \omega
_0\right)}$,\\
$\frac{\sqrt{\frac{2}{3}} (\beta -2) \beta \lambda }{2 \beta ^2+(\beta +1)
\lambda ^2+2 \lambda ^2 \omega _0}, \lambda\Big)$.
\end{widetext}
The eigenvalues and the nature of the equilibrium points has to be handled for
specific choices of the parameters in the region of existence.
\begin{widetext}
\item[$P_{12}(\lambda)$] $:=(x_1,y,z,\lambda)=\Big( 0, \frac{\sqrt{2 \lambda ^2 \omega _0 \left((14-(\beta -4) \beta )
\lambda ^2+6 \beta (\beta +2)+6 \lambda ^2 \omega _0\right)-3 \left((\beta
-5) \lambda ^2-4 \beta \right) \left(2 \beta ^2+(\beta +1) \lambda ^2\right)}%
}{\sqrt{3} \left(2 \beta ^2+(\beta +1) \lambda ^2+2 \lambda ^2 \omega
_0\right)}$,\\
$\frac{\sqrt{\frac{2}{3}} (\beta -2) \beta \lambda }{2 \beta ^2+(\beta +1)
\lambda ^2+2 \lambda ^2 \omega _0}, \lambda\Big)$.
\end{widetext}
The eigenvalues and the nature of the equilibrium points have to be handled for
specific choices of the parameters in the region of existence.

\item[$P_{13}$] $:=(x_1,y,z,\lambda)=\left(0,0,\sqrt{-\frac{2 \omega _0}{3}-1},%
\frac{\beta }{\sqrt{-\omega _0-\frac{3}{2}}}\right)$. \newline
The eigenvalues are 
$\{3-\beta ,1,-2 (\beta -3),1\}$.\newline
It is a source for

\begin{enumerate}
\item $\omega _0\in \mathbb{R}, \beta =0$, or

\item $0<\beta <3, \omega  _0<-\frac{3}{2}$.
\end{enumerate}

\begin{widetext}
\item[$P_{14}$] $:= (x_1,y,z,\lambda)= {\left(0,0,\frac{1}{96 \beta ^2 \sqrt{%
(-10+2 \beta ) (2+2 \beta )-24 \omega _0} \omega _0 \sqrt{3+2 \omega _0}}%
\right.}\newline
{\left(-48 \sqrt{3} \omega _0^2+3 (2+2 \beta ) \left(-2 \sqrt{3}+2 \sqrt{3}
\beta +\sqrt{3 (-2+2 \beta )^2+8 \omega _0 \left(6+2 \beta +6 \omega
_0\right)}\right)+\right.}\newline
{\left.4 \omega _0 \left(\sqrt{3} (-6+2 \beta ) (2+2 \beta )+3 \sqrt{3 (-2+2
\beta )^2+8 \omega _0 \left(6+2 \beta +6 \omega _0\right)}\right)\right) }%
\newline
{\surd (-6 \beta (-8+2 \beta (-14+2 \beta ))+}\newline
{\left.2 \beta \left(12 (4+2 \beta ) \omega _0-(-4+2 \beta ) \sqrt{9 (-2+2
\beta )^2+24 \omega _0 \left(6+2 \beta +6 \omega _0\right)}\right)\right),}%
\newline
{\left.\left(8 \sqrt{3} \beta ^{3/2}\right)\right/(\surd (6 \beta (-8+2
\beta (-14+2 \beta ))-}\newline
{\left.\left.\left.2 \beta \left(12 (4+2 \beta ) \omega _0+(-4+2 \beta )
\sqrt{9 (-2+2 \beta )^2+24 \omega _0 \left(6+2 \beta +6 \omega _0\right)}%
\right)\right)\right)\right)}$.
\end{widetext}
The eigenvalues and the nature of the equilibrium points have to be handled for
specific choices of the parameters in the region of existence.
\begin{widetext}
\item[$P_{15}$] $:=(x_1,y,z,\lambda)={\left(0,0,\frac{1}{96 \beta ^2 \sqrt{%
(-10+2 \beta ) (2+2 \beta )-24 \omega _0} \omega _0 \sqrt{3+2 \omega _0}}%
\right.}\newline
{\left(-48 \sqrt{3} \omega _0^2+4 \omega _0 \left(\sqrt{3} (-6+2 \beta )
(2+2 \beta )-3 \sqrt{3 (-2+2 \beta )^2+8 \omega _0 \left(6+2 \beta +6 \omega
_0\right)}\right)+\right.}\newline
{\left.3 (2+2 \beta ) \left(-2 \sqrt{3}+2 \sqrt{3} \beta -\sqrt{3 (-2+2
\beta )^2+8 \omega _0 \left(6+2 \beta +6 \omega _0\right)}\right)\right) }%
\newline
{\surd (-6 \beta (-8+2 \beta (-14+2 \beta ))+}\newline
{\left.2 \beta \left(12 (4+2 \beta ) \omega _0+(-4+2 \beta ) \sqrt{9 (-2+2
\beta )^2+24 \omega _0 \left(6+2 \beta +6 \omega _0\right)}\right)\right),}%
\newline
{\left.\left(8 \sqrt{3} \beta ^{3/2}\right)\right/(\surd (6 \beta (-8+2
\beta (-14+2 \beta ))+}\newline
{\left.\left.\left.2 \beta \left(-12 (4+2 \beta ) \omega _0+(-4+2 \beta )
\sqrt{9 (-2+2 \beta )^2+24 \omega _0 \left(6+2 \beta +6 \omega _0\right)}%
\right)\right)\right)\right)}$.
\end{widetext}
The eigenvalues and the nature of the equilibrium points have to be handled for
specific choices of the parameters in the region of existence.
\end{enumerate}

\subsubsection{Center manifold of $P_1$.}
\label{Center_manifolds_P1_b}

From the previous linear analysis we found that the equilibrium point $P_1$ is nonhyperbolic with a three dimensional stable manifold provided
$\beta >2, \omega _0\geq \frac{1%
}{6} (16 \beta -41)$.

Introducing the new variables
\begin{align*}
& u=\lambda, \\
& v_1= z-\frac{(\beta -2) \lambda }{\sqrt{6} \beta } \\
& v_2=-\frac{x \left(4 (\beta -5) \omega _0+6 (\beta -1)\right)}{2 \sqrt{\beta } \sqrt{\left(2 \omega _0+3\right) \left(-16 \beta +6 \omega _0+41\right)}} \nonumber \\
& + 
 \frac{\left(\sqrt{\left(2 \omega _0+3\right) \left(-16 \beta +6 \omega _0+41\right)}-2 \sqrt{3} \omega _0+5 \sqrt{3}\right) \left(y-\frac{\sqrt{2}}{\sqrt{\beta }}\right)}{2 \sqrt{\left(2 \omega _0+3\right) \left(-16 \beta
   +6 \omega _0+41\right)}},\\
& v_3=\frac{x \left(4 (\beta -5) \omega _0+6 (\beta -1)\right)}{2 \sqrt{\beta } \sqrt{\left(2 \omega _0+3\right) \left(-16 \beta +6 \omega _0+41\right)}}
\nonumber \\
& +\frac{\left(\sqrt{\left(2 \omega _0+3\right) \left(-16 \beta +6 \omega
   _0+41\right)}+2 \sqrt{3} \omega _0-5 \sqrt{3}\right) \left(y-\frac{\sqrt{2}}{\sqrt{\beta }}\right)}{2 \sqrt{\left(2 \omega _0+3\right) \left(-16 \beta +6 \omega _0+41\right)}},
\end{align*}

we obtain the evolution equations
\begin{subequations}
\label{JBDcenter2B}
\begin{align}
&u'=-\frac{\sqrt{\beta
   } u v_2 \left(\sqrt{3} \sqrt{(2 \omega_0+3) (-16 \beta +6 \omega_0+41)}+6 \omega_0-15\right)}{2 \sqrt{2} (2 (\beta -5) \omega_0+3 (\beta -1))}+\nonumber \\
	& \frac{\sqrt{\beta } u v_3 \left(\sqrt{3} \sqrt{(2 \omega_0+3) (-16 \beta +6 \omega_0+41)}-6 \omega_0+15\right)}{2 \sqrt{2} (2 (\beta -5) \omega_0+3 (\beta -1))},\\
&v_1'=	-3 v_1 +\mathcal{O}(2),\\
&v_2'=-\frac{v_2 \left(\sqrt{3} \sqrt{(2 \omega_0+3) (-16 \beta +6
   \omega_0+41)}+6 \omega_0+9\right)}{4 \omega_0+6}+\mathcal{O}(2),\\
& v_3'=	\frac{v_3 \left(\sqrt{3} \sqrt{(2 \omega_0+3) (-16 \beta +6 \omega_0+41)}-6 \omega_0-9\right)}{4
   \omega_0+6}+\mathcal{O}(2).
\end{align}
\end{subequations}
The system \eqref{JBDcenter2B} can be diagonalized as
\begin{align}
u'  &  =Cu+f\left(  u,\mathbf{v}\right),\nonumber\\
\mathbf{v}'  &  =P\mathbf{v}+\mathbf{g}\left(  u,\mathbf{v}\right), \label{JBDcenter3B}
\end{align}
where $\left(  u,\mathbf{v}\right)
\in\mathbb{R}\times\mathbb{R}^{3},$ $C\equiv 0 \in \mathbb{R}^{1\times 1}$, $P \in \mathbb{R}^{3\times 3}$ has negative eigenvalues, and  $f(0)=0, \quad \mathbf{g}(\mathbf{0})=\mathbf{0}$, $f'(0)=0, \quad D\mathbf{g}(\mathbf{0})=\mathbf{0}$. Using the center manifold theorem we have that there exists a 1-dimensional
invariant local center manifold $W^{c}\left( \mathbf{0}\right) $
of \eqref{JBDcenter3B} tangent to the center subspace 
$\mathbf{v}=\mathbf{0}$ at $\mathbf{0}.$ Moreover,
$W^{c}\left( \mathbf{0}\right)  $ can be represented as
\begin{align}
& W^{c}\left(  \mathbf{0}\right)  =\Big\{  \left(
u,\mathbf{v}\right)
\in\mathbb{R}\times\mathbb{R}^{3}:\mathbf{v}=\mathbf{h}\left(
u\right), \mathbf{h}\left(  0\right) =\mathbf{0},  D\mathbf{h}\left(
0\right)  =\mathbf{0},\;\left\vert u\right\vert <\delta\Big\},
\end{align}
for $\delta$ small enough. The dynamics over the center manifold is given by the equation
\begin{equation}
u'=f\left( u,\mathbf{h}\left(  u\right)  \right),
\end{equation}
where the function $\mathbf{h}\left( u\right)  $ defines the local
center manifold and
satisfies%
\[D\mathbf{h}\left(  u\right)  \left[  f\left(  u,\mathbf{h}\left(
u\right) \right)  \right]  -P\mathbf{h}\left(  u\right)
-\mathbf{g}\left( u,\mathbf{h}\left(  u\right)  \right)  =0.
\]
Following the same implemented procedure in section \ref{Center_manifolds_P1_a}
we obtain 
\[
\mathbf{h}\left(  u\right)  :=\left[
\begin{array}
[c]{c}%
h_{1}\left( u\right) \\
h_{2}\left(  u\right) \\
h_{3}\left(  u\right)
\end{array}
\right]  =\left[
\begin{array}
[c]{c}%
a_{1}u^{2}+\mathcal{O}\left(  u^{3}\right) \\
a_{2}u^{2}+\mathcal{O}\left(  u^{3}\right) \\
a_{3}u^{2}+\mathcal{O}\left(  u^{3}\right)
\end{array}
\right]
\]
\begin{widetext}
where $a_1=0$,\\
\[a_2= \frac{\sqrt{2} (\beta -2)^2 \left(\omega_0 \sqrt{(2 \omega_0+3) (-16 \beta +6 \omega_0+41)}+\sqrt{3} \beta  (2 \omega_0+3)-\sqrt{3} (\omega_0+1) (2
   \omega_0+3)\right)}{\beta ^{5/2} (2 \omega_0+3) \left(3 \sqrt{(2 \omega_0+3) (-(16 \beta -6 \omega_0-41))}-16 \sqrt{3} \beta +\sqrt{3} (6 \omega_0+41)\right)},\]
\[a_3=
   \frac{\sqrt{2} (\beta -2)^2 \left(\omega_0 \left(-\sqrt{(2 \omega_0+3) (-16 \beta +6 \omega_0+41)}\right)+\sqrt{3} \beta  (2 \omega_0+3)-\sqrt{3} (\omega_0+1) (2
   \omega_0+3)\right)}{\beta ^{5/2} (2 \omega_0+3) \left(-3 \sqrt{(2 \omega_0+3) (-(16 \beta -6 \omega_0-41))}-16 \sqrt{3} \beta +\sqrt{3} (6 \omega_0+41)\right)}.\]
	\end{widetext}
Therefore, the dynamics on the center manifold is given by the gradient-like equation 
\[u'=-\frac{(\beta -2) u^3}{2 \beta ^2}=-\frac{d}{d u} \left[\frac{(\beta -2) u^4}{8 \beta ^2}\right],\]
under the potential $\Pi(u)=\frac{(\beta -2) u^4}{8 \beta ^2}$, for which the origin is a degenerate local minimum whenever $\beta>2$ (recall the existence conditions for $P_1$ are $\beta >2, \omega _0\geq \frac{1}{6} (16 \beta -41)$), and under these conditions, the center manifold of $P_1$ is stable.
In the original variables  \eqref{coordinates-x-y-z} the center manifold can be locally expressed as the graph 
\begin{subequations}
\begin{align}
& x= -\frac{(\beta -2) u^2}{\sqrt{6} \beta ^2}, \\
& y= -\frac{(\beta -2) u^2 (\beta -2 \omega_{0}-1)-8 \beta ^2}{4 \sqrt{2} \beta ^{5/2}}, \\
& z=\frac{(\beta -2) u}{\sqrt{6} \beta }, \\
& \lambda= u.
\end{align}
\end{subequations}
According to the center manifold calculation, we have from \eqref{Center_coordinates}, the definition $\lambda:=-\frac{2 \beta}{\psi} e^{-\varphi/2}$, and the definition \eqref{coordinates-x-y-z}, and introducing the time rescaling $\frac{d f}{d \tau}=\psi^2 \frac{d f}{d \ln a}$ we have
that (as $\lambda\rightarrow 0$):
\begin{subequations} 
\begin{align}
& \frac{d \varphi}{d \tau}={4 (2-\beta) e^{-\varphi }}, \\
& \frac{d \psi}{d \tau}={2(2- \beta)}{\psi }, \\
&\frac{d t}{d \tau}=\frac{\sqrt{\frac{3}{2}} e^{-\frac{1}{2} (\beta +1) \varphi } \left(2 \psi ^2 e^{\phi }-(\beta -2) (\beta -2 \omega_{0}-1)\right)}{\sqrt{\beta }
   \sqrt{V_0}}.
\end{align}
\end{subequations}
Integrating the equations, and using the first integral $\ln\left[\frac{a}{a_0}\right]=\int \psi^2 d\tau$  we have the general solution
\begin{subequations}
\begin{align}
& \varphi (\tau )=\ln\left|c_1-4 (\beta -2) \tau \right|, \\
&\psi (\tau )= c_2 e^{-2 (\beta -2) \tau }, \\
& a=a_0\exp\left[-\frac{c_2^2 e^{-4 (\beta -2) \tau }}{4 (\beta -2)}\right],
\end{align}
\begin{widetext}
\begin{align}
&t-t_0=\frac{\sqrt{\frac{3}{2}} \int \left|c_1-4 (\beta -2) \tau \right|{}^{\frac{1}{2} (-\beta -1)} \left((2-\beta ) (\beta -2 \omega_{0}-1)+2 c_2^2 e^{-4 (\beta -2) \tau } \left|c_1-4 (\beta -2) \tau \right|\right) \, d\tau }{\sqrt{\beta } \sqrt{V_{0}}}\nonumber \\
&\simeq -\frac{\sqrt{\frac{3}{2}} (\beta -2 \omega_{0}-1) \left|-4 \beta  \tau +c_1+8 \tau \right|{}^{\frac{1}{2}-\frac{\beta }{2}}}{2 (\beta -1) \sqrt{\beta } \sqrt{V_{0}}} \; \text{for large}\; \tau.
\end{align}
\end{widetext}
\end{subequations}

\section{Cosmological consequences}
\label{Sect:5}

Considering the equations written in the dilatonic frame: (\ref{fe.01}),(\ref{fe.02}), (\ref{fe.03}), and (\ref{fe.04}), we can define the following observable quantities:
\begin{subequations}
\begin{align}
&  {\Omega_1}\equiv e^{-\varphi}\frac{ {\rho_1}}{3 H^2},\;%
\text{where}\;  {\rho_1}:= e^{\varphi }\left( \frac{\omega _{0}\dot{
\varphi}^{2}}{2}-3H\dot{\varphi}+V(\varphi )\right), \\
& {\Omega_2}\equiv e^{-\varphi}\frac{\rho_2}{3 H^2},\;%
\text{where}\;  {\rho_2}:=\frac{1}{2}{\dot{\psi}}^{2}+W(\psi ),\\
& {w_1}\equiv \frac{ {p_1}}{ {\rho_1}}, \; \text{%
where}\;  {p_1}:=\frac{4 \left(3-2 \omega _0\right) H e^{\varphi}
\dot\varphi}{4 \omega _0+6}+\frac{12 H^2 e^{\varphi}}{2 \omega _0+3} \nonumber \\
&   -\frac{e^{\varphi} \left(4 V^{\prime }(\varphi)+2 \left(2 \omega
_0+5\right) V(\varphi)+\left(1-2 \omega _0\right) \omega _0
\dot\phi^2\right)+6 \dot\psi^2}{4 \omega _0+6}, \\
& w_2=w_\psi \equiv \frac{p_2}{\rho_2}
 \label{Eq49}, \; \text{%
where}\;   p_2:=\frac{1}{2}{\dot{\psi}}^{2}-W(\psi ), \;\text{and}\;\\
& {w_{\text{tot}}}:= \frac{ {p_1}+ {p_2}}{%
 {\rho_1}+ {\rho_2}}.
\end{align}
\end{subequations}
These cosmological parameters can be written in terms of the phase space variables
expressed as
\begin{subequations}
\begin{align}
& {\Omega_1}=x^2 \omega _0-\sqrt{6} x+y^2, \\
& {\Omega_2}=-x^2\omega _0+\sqrt{6} x-y^2+1,  \label{Eq69}
\\
& {w_1}=\frac{3 \left((2 s-5) y^2-2 \omega _0 y^2-6 z^2+4\right)+3
x^2 \omega _0 \left(2 \omega _0-1\right)+2 \sqrt{6} x \left(3-2 \omega
_0\right)}{3 \left(2 \omega _0+3\right) \left(x^2 \omega _0-\sqrt{6}
x+y^2\right)},  \label{Eq70}
\\
& {w_2}=\frac{x^2 \omega _0-\sqrt{6} x+y^2+2 z^2-1}{x^2
\left(-\omega _0\right)+\sqrt{6} x-y^2+1},  \label{Eq71} \\
& {w_{\text{tot}}}=\frac{6 (s-1) y^2+2 \omega _0 \left(6 x^2 \omega
_0+3 x^2-5 \sqrt{6} x+6 z^2-3\right)-3 \sqrt{6} x+3}{6 \omega _0+9}.
\label{Eq72}
\end{align}
\end{subequations}

$w_{\text{tot}}$ is
related to the deceleration for isotropic metrics by $q=\frac{1}{2}(1+3w_{\text{tot}})$.

We continue with the discussion for the
interpretations of the model for the choices studied in sections: \ref%
{sect:4.1}, and \ref{sect:4.2}. We finish the section with a discussion of
the generic features of the models.

\subsection{Case: $W(\protect\psi)=W_0 e^{k \protect\psi}$ and $V(\protect%
\varphi)=V_0 e^{l \protect\varphi}$}
\label{Sect:6.1}

\begin{table*}
\caption{Cosmological parameters corresponding to the formulation in the
dilatonic frame as given by Eqs. \eqref{FE_jordan} for $W(\protect\psi)=W_0
e^{k \protect\psi}$ and $V(\protect\varphi)=V_0 e^{l \protect\varphi}$.}
\label{Tab2b}
\begin{tabular*}{\textwidth}{@{\extracolsep{\fill}}lrrrl@{}}
\hline
Equil. & \multicolumn{1}{c}{$\Omega_2$} & \multicolumn{1}{c}{$w_1$} & \multicolumn{1}{c}{$w_2$} & \multicolumn{1}{c}{${w_{\text{tot}}}$}  \\
Points \\
\hline
$P_1$  & $\frac{(l-1)}{l+1}$ & $-1$ & $-1$ & $-1$ \\ \hline
$P_2$ & $-\frac{2 \omega _0}{3}-1$ & $1$ & $1$ & $1$  \\
\hline
$P_3$  & $\frac{ \left((l-3) l-6 \omega _0-7\right)}{(l+1)^2}$& $1$ & $1$  & $1$ \\\hline
   $P_4$ &$\frac{ \left((l-3) l-6 \omega _0-7\right)}{(l+1)^2}$ & $1$ & $1$  & $1$   \\\hline
   $P_5$ & $\frac{\left(2 \omega _0+3\right) \left(6 \omega _0+5\right)}{3 \left(2 \omega _0+1\right){}^2}$ & $-1$ & $-1$ & $-1$ \\\hline
$P_6$ & $0$ & $\frac{2 \left(l^2-1\right)}{3 \left(l+2 \omega _0+2\right)}-1$ & Indeterminate & $\frac{2 \left(l^2-1\right)}{3 \left(l+2 \omega _0+2\right)}-1$\\\hline
$P_7$ & $0$ & $\frac{2 \left(l^2-1\right)}{3 \left(l+2 \omega _0+2\right)}-1$ & Indeterminate & $\frac{2 \left(l^2-1\right)}{3 \left(l+2 \omega _0+2\right)}-1$\\\hline
$P_8$ & $0$ & $\frac{3 \omega _0-\sqrt{6 \omega _0+9}+3}{3 \omega _0}$ & $-1$ & $\frac{3 \omega _0-\sqrt{6 \omega _0+9}+3}{3 \omega _0}$\\\hline
$P_9$ & $0$ & $\frac{3 \omega _0+\sqrt{6 \omega _0+9}+3}{3 \omega _0}$ & $-1$ & $\frac{3 \omega _0+\sqrt{6 \omega _0+9}+3}{3 \omega _0}$\\\hline
$P_{10}$ & $1$ & \text{Indeterminate}& $\frac{1}{3}$ & $\frac{1}{3}$\\
\hline
\end{tabular*}
\end{table*}
In Table \ref%
{Tab2b} we present the cosmological parameters corresponding to the
formulation in the dilatonic frame as given by Eqs. \eqref{FE_jordan} for $W(\psi)=W_0 e^{k \psi}$ and $V(\varphi)=V_0 e^{l
\varphi}$. We have the following results:
\begin{itemize}
\item $P_1$ satisfies  $\Omega_2=1-\Omega_1=\frac{(l-1)}{l+1}$ with $w_1=-1$, $w_2=-1$ and ${w_{\text{tot}}}=-1$. 
Both energy densities $\Omega_1, \Omega_2$ are in the same order of magnitude, that is a scaling solution. We have proved that its center manifold is stable for $1<l\leq \frac{1}{16} \left(6 \omega _0+25\right)$. Hence, this point is a late-time attractor.

\item The equilibrium points $P_2, P_3$ and $P_4$ satisfy $w_1=1$,  $w_2=1$,  ${w_{\text{tot}}}=1$. That is, they represent stiff solutions. The three solutions are saddle therefore they are not relevant for the late-time cosmology, neither for the early-time cosmology.

\item The equilibrium point $P_5$, which exists $\omega _0<-\frac{3}{2}$ or  $-\frac{5}{6}\leq \omega  _0<-\frac{1}{2}$ or $\omega _0>-\frac{1}{2}$,  satisfies $\Omega_2=\frac{\left(2 \omega _0+3\right) \left(6 \omega _0+5\right)}{3 \left(2 \omega _0+1\right){}^2}$, $w_1=-1$, $w_2=-1$ and ${w_{\text{tot}}}=-1$. 
Both energy densities $\Omega_1, \Omega_2$ are of the same order of magnitude, that is a scaling solution.  It is a sink for $l>-1, \omega _0<-\frac{3}{2}$ or a saddle otherwise.

\item The equilibrium point $P_6$,  which exists for  $l<-2, \omega _0=\frac{1}{6} (l-4) (l+2)$ or $l>1, \omega  _0=\frac{1}{6} (l-4) (l+2)$ or  $-2<l<1, \frac{1}{6} (l-4) (l+2)\leq  \omega _0<\frac{1}{2} (-l-2)$ or $l\leq 1, \omega  _0<-\frac{3}{2}$ or  $l>1, \omega _0<\frac{1}{2} (-l-2)$, satisfies $\Omega_2=0$, that is, the energy density of the dilatonic field is dominant and the energy density of the quintessence field is negligible. Furthermore,  $w_1=\frac{2 \left(l^2-1\right)}{3 \left(l+2 \omega _0+2\right)}-1$ and ${w_{\text{tot}}}=\frac{2 \left(l^2-1\right)}{3 \left(l+2 \omega _0+2\right)}-1$. That is, according to whether ${w_{\text{tot}}}<-1, {w_{\text{tot}}}=-1$ or $-1<{w_{\text{tot}}}<-\frac{1}{3}$ it represents a phantom solution, a solution with $w_{\text{tot}}=-1$ or a quintessence solution.  It is a sink for $l<-1, \omega _0<-\frac{3}{2}$. It is a saddle otherwise.

\item The equilibrium point $P_7$ exists for $l>1, \omega _0\geq \frac{1}{6}(l-4) (l+2)$ or  $l<-2,  \omega _0\geq \frac{1}{6} (l-4) (l+2)$ or $l=-2, \omega  _0>0$ or $l=1, \omega _0>-\frac{3}{2}$ or  $-2<l<1, \omega  _0>\frac{1}{2} (-l-2)$ or  $l>1, \frac{1}{2} (-l-2)<\omega  _0<-\frac{3}{2}$ or  $-2<l<1, \omega _0=\frac{1}{6} (l-4) (l+2)$. The cosmological observables are $\Omega_2=0$, $w_1=\frac{2 \left(l^2-1\right)}{3 \left(l+2 \omega _0+2\right)}-1$ and ${w_{\text{tot}}}=\frac{2 \left(l^2-1\right)}{3 \left(l+2 \omega _0+2\right)}-1$. It is a saddle,  therefore they are not relevant for the late-time cosmology, neither for the early-time cosmology.

\item The equilibrium point $P_8$, which exists for $\omega _0>-\frac{3}{2}, \omega _0\neq 0$, satisfies   $\Omega_2=0$ that is, the energy density of the dilatonic field is dominant and the energy density of the quintessence field is negligible. Furthermore,\\$w_1=\frac{3 \omega _0-\sqrt{6 \omega _0+9}+3}{3 \omega _0}$, $w_2=-1$, and $w_{\text{tot}}=\frac{3 \omega _0-\sqrt{6 \omega _0+9}+3}{3 \omega _0}$. The total energy density represents a standard matter source with $0<w_{\text{tot}}<1$.
It is a saddle. Therefore, they are not relevant for the late-time cosmology neither for the early-time cosmology.

\item The equilibrium point $P_9$, which exists for $\omega _0>-\frac{3}{2}, \omega _0\neq 0$, satisfies   $\Omega_2=0$ that is, the energy density of the dilatonic field is dominant and the energy density of the quintessence field is negligible. Furthermore, \\$w_1=\frac{3 \omega _0+\sqrt{6 \omega _0+9}+3}{3 \omega _0}$, $w_2=-1$, and $w_{\text{tot}}=\frac{3 \omega _0+\sqrt{6 \omega _0+9}+3}{3 \omega _0}$. That is, the second fluid behaves as a cosmological constant whereas the effective equation of state (of the total cosmic budget) is that of quintessence field for $-\frac{9}{8}<\omega_0<-\frac{5}{6}$,  a cosmological constant for $\omega_0=-\frac{5}{6}$ and the phantom field for $-\frac{5}{6}<\omega_0<0$.
It is a sink for $-2<l\leq -1, \frac{1}{6} (l-4) (l+2)<\omega _0<0$, or  $l>-1, -\frac{5}{6}<\omega _0<0$ (in both cases it is a phantom attractor). It is a source for $l\leq -2, \omega _0>\frac{1}{6} (l-4) (l+2)$, or  $l>-2, \omega_0>0$ (and then, it behaves as a standard matter source ).
It is a saddle otherwise.

\item The equilibrium point $P_{10}$ satisfies $\Omega_2=1$. That is dominated by the quintessence field and the contribution of the dilatonic field to the total energy density is negligible. It satisfies $w_2=\frac{1}{3}$ and ${w_{\text{tot}}}=\frac{1}{3}$. This means that the corresponding cosmological solution mimics radiation. Interestingly, it is a saddle with a three dimensional stable manifold provided $\omega_0\geq \frac{45}{2}$.

\end{itemize}

\begin{figure}[ht]
{\includegraphics[width=0.6\textwidth]{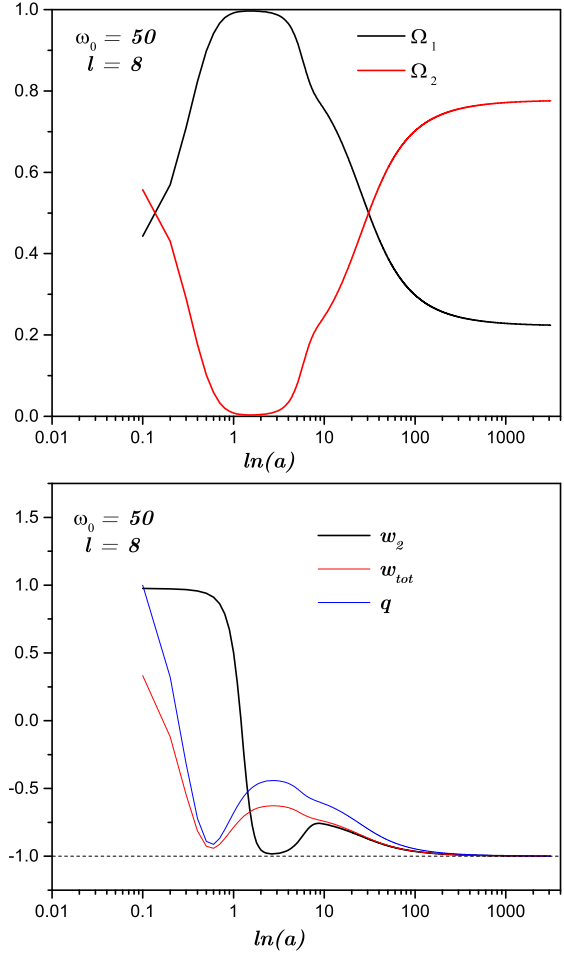}}
\caption{\label{FIG1b} Qualitative evolution of the dimensionless energy densities $\Omega_1, \Omega_2$ and the observables $w_2, {w_{\text{tot}}}, q$ vs $\ln(a)$  for $W(\protect\psi)=W_0 e^{k \protect\psi}$ and $V(\protect%
\varphi)=V_0 e^{l \protect\varphi}$ with $\omega_0=50, l=8$. }
\end{figure}

In the Fig. \ref{FIG1b} is presented evolution of the dimensionless energy densities $\Omega_1, \Omega_2$ and the observables $w_2, {w_{\text{tot}}}, q$ vs $\ln(a)$  for $W(\protect\psi)=W_0 e^{k \protect\psi}$ and $V(\protect%
\varphi)=V_0 e^{l \protect\varphi}$ with $\omega_0=50, l=8$.

\begin{figure}[ht]
{\includegraphics[width=0.6\textwidth]{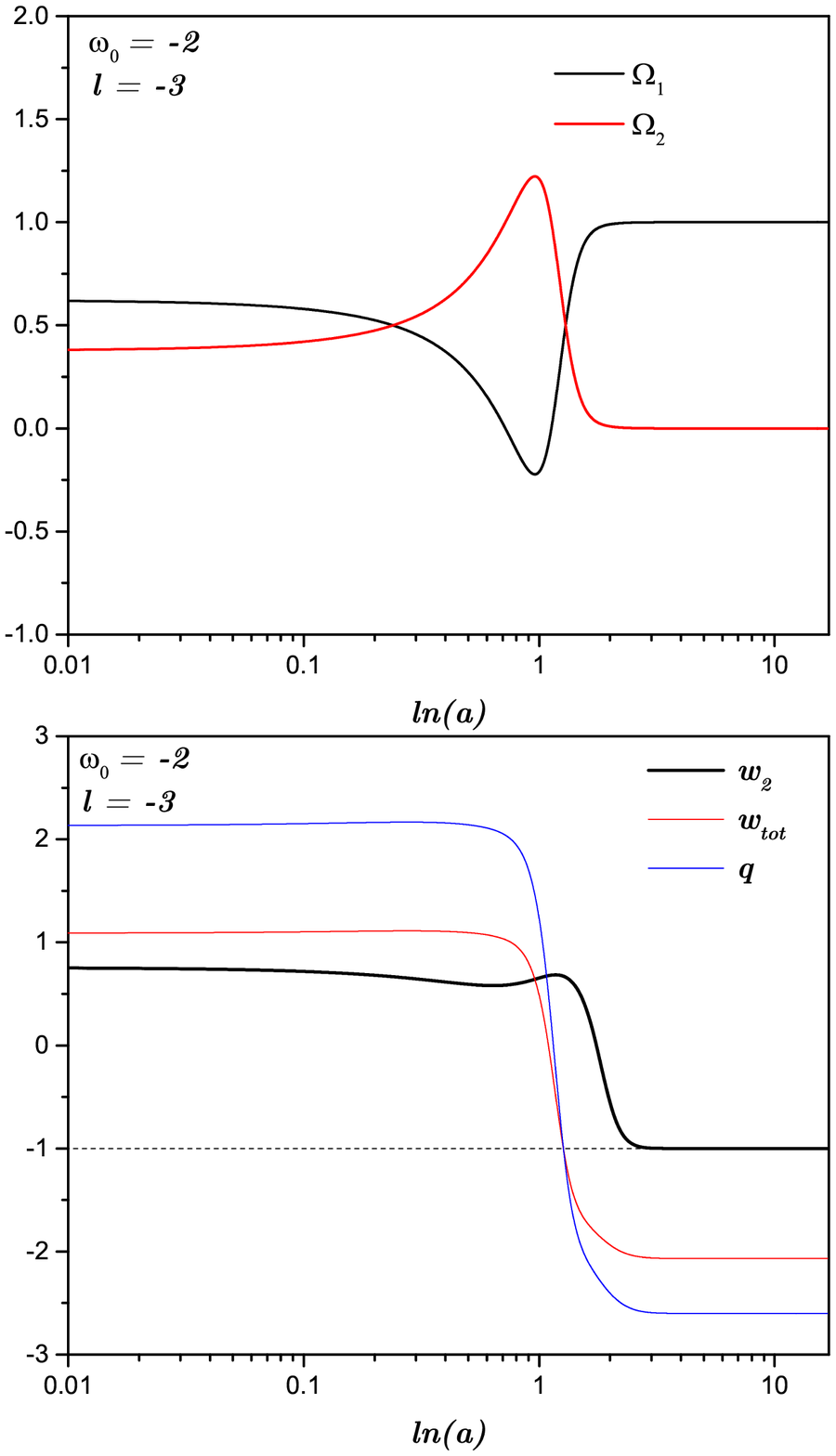}}
\caption{\label{FIG2b} Qualitative evolution of the dimensionless energy densities $\Omega_1, \Omega_2$ and the observables $w_2, {w_{\text{tot}}}, q$ vs $\ln(a)$  for $W(\protect\psi)=W_0 e^{k \protect\psi}$ and $V(\protect%
\varphi)=V_0 e^{l \protect\varphi}$ with $\omega_0=-2, l=-3$. }
\end{figure}

In the Fig. \ref{FIG2b} is presented evolution of the dimensionless energy densities $\Omega_1, \Omega_2$ and the observables $w_2, {w_{\text{tot}}}, q$ vs $\ln(a)$  for $W(\protect\psi)=W_0 e^{k \protect\psi}$ and $V(\protect%
\varphi)=V_0 e^{l \protect\varphi}$ with $\omega_0=-2, l=-3$.

\subsection{Case: $W(\protect\psi)=W_0 \protect\psi^{2\protect\beta}$ and $V(%
\protect\varphi)=V_0 e^{(\protect\beta-1) \protect\varphi}$}

For this model we have the same results for the first nine equilibrium
points in Table \ref{Tab2b}, in section \ref%
{Sect:6.1} replacing $l=\beta-1, k=2\beta$, and the additional equilibrium
points $P_{11}-P_{15}$.

The relevant early- and late-time attractors are the following:
\begin{itemize}
\item  The stable manifold of $P_1$ is three dimensional for $\beta >2, \omega _0\geq \frac{1}{6} (16 \beta -41)$.
\item The equilibrium point $P_5$ is a sink for $\beta >0, \omega _0<-\frac{3}{2}$.
\item The equilibrium point $P_6$ is a sink for $\beta <0, \omega _0<-\frac{3}{2}$.
\item The equilibrium point $P_6$  is a sink for $-1<\beta \leq 0, \frac{1}{6} (\beta -5) (\beta +1)<\omega _0<0$, or
 $\beta >0, -\frac{5}{6}<\omega _0<0$. It is a source for $\beta \leq -1, \omega _0>\frac{1}{6} (\beta -5) (\beta +1)$, or
 $\beta  >-1, \omega _0>0$.
\item $P_{13}$  is a source for  $\omega _0\in \mathbb{R}, \beta =0$, or  $0<\beta <3, \omega  _0<-\frac{3}{2}$.
\item  The observables for $P_{11, 12}$ and $P_{14, 15}$ have to be evaluated for specific choices of the parameters.

\end{itemize}

\section{Some results for arbitrary potentials}

\label{Sect:6}  In the sections \ref{sect:4.1} and \ref{sect:4.2}, we have
investigated an exponential potential $V(\varphi)$ for which $s$ is a
constant such that $h$ is only a function of $\lambda$ that depends on the
choice of $W$, and $F$ is identically zero. For complementing these results, in
this section we comment about the generic features of the equilibrium points of %
\eqref{syst1} for arbitrary $h(\lambda, s)$ and $F(s)$ functions.

Since the system is form-invariant under the change\\$(y,
z,\lambda)\rightarrow (-y,-z, -\lambda)$. Thus, without losing generality we
can investigate just the sector $y\geq 0, z\geq 0, \lambda\geq 0$.
Henceforth, we will focus on the stability properties of the system %
\eqref{syst1} for the state vector $(x,y,z, \lambda)$ defined in the phase
space
\begin{align}
&\Big\{(x,y,z, s, \lambda)\in\mathbb{R}^5: -\sqrt{6} x + x^2 \omega_0 +y^2 +z^2 \leq 1, y\geq 0, z\geq 0, \lambda\geq 0, s\in\mathbb{R} \Big\}.
\end{align}

The equilibrium points of \eqref{syst1} that are independent of $h(s,\lambda)
$ are summarized below.
In table \ref{Tab3b} are shown
the cosmological parameters corresponding to the formulation in the
dilatonic frame as given by Eqs. \eqref{FE_jordan} for the equilibrium points in the invariant set $\lambda=0$ for arbitrary
potentials.

\begin{enumerate}
\item[$P_1$:] $(x,y,z,\lambda,s)=\left(0, \frac{\sqrt{2}}{\sqrt{1-s_c}}%
,0,0,s_c\right)$. \newline
Exists for $s_c\leq -1$. \newline
The eigenvalues are \newline
$\Big\{0,0,-3$,\\
$\frac{1}{2} \left(-\frac{\sqrt{3} \sqrt{\left(s_c-1\right)
\left(16 F\left(s_c\right)+\left(s_c-1\right) \left(16 s_c+6 \omega
_0+25\right)\right)}}{\sqrt{2 \omega _0+3} \left(s_c-1\right)}-3\right)$,%
\newline
$\frac{1}{2} \left(\frac{\sqrt{3} \sqrt{\left(s_c-1\right) \left(16
F\left(s_c\right)+\left(s_c-1\right) \left(16 s_c+6 \omega
_0+25\right)\right)}}{\sqrt{2 \omega _0+3} \left(s_c-1\right)}-3\right)\Big\}
$.\newline
This line of equilibrium points contains the cases $s_c=\hat{s}: F(\hat{s})=0$,
for which the eigenvalues simplifies to \newline
$\left\{0,0,-3,\frac{1}{2} \left(-\frac{\sqrt{16 \hat{s}+6 \omega _0+25}}{%
\sqrt{2 \omega _0+3}}-3\right),\frac{1}{2} \left(\frac{\sqrt{16 \hat{s}+6
\omega _0+25}}{\sqrt{2 \omega _0+3}}-3\right)\right\}$. \newline
It is non-hyperbolic. The stable manifold of $P_1$ is three dimensional
provided\\$s_c\leq -1, 1-s_c^2<F\left(s_c\right)\leq -\frac{1}{16}
\left(s_c-1\right) \left(16 s_c+6 \omega _0+25\right)$, in other case the
stable manifold is a lower dimensional.

\item[$P_2(\hat{s})$:] $(x,y,z,\lambda,s)=\left(-\sqrt{\frac{2}{3}},0,\sqrt{-%
\frac{2 \omega _0}{3}-1},0,\hat{s}\right)$, where $F(\hat{s})=0$. \newline
Exists for $\omega _0<-\frac{3}{2}$. \newline
The eigenvalues are  $\left\{6,-1,1,\hat{s}+2,2 F^{\prime }\left(\hat{s}%
\right)\right\}$. \newline
The equilibrium point is a saddle and has a four dimensional unstable
manifold provided $\hat{s}>-2, F^{\prime }(\hat{s})>0$. In other case, the
unstable manifold is a lower dimensional.

\item[$P_3(\hat{s})$:] $(x,y,z,\lambda,s)=\left(\frac{\sqrt{6}}{\hat{s}-1},%
\frac{\sqrt{\hat{s}+2}}{\hat{s}-1},\frac{\sqrt{\hat{s} \left(\hat{s}%
+3\right)-6 \omega _0-7}}{\hat{s}-1},0,\hat{s}\right)$. \newline
Exists for

\begin{enumerate}
\item $\hat{s}>1, \omega _0<-\frac{3}{2}$, or

\item $\hat{s}>1, -\frac{3}{2}<\omega _0\leq \frac{1}{6} \left(\hat{s}^2+3
\hat{s}-7\right)$.
\end{enumerate}

The eigenvalues are\newline
$\Big\{\frac{3}{\hat{s}-1},6,-\frac{3}{2 \left(\hat{s}-1\right)}-\frac{\sqrt{%
-8 \hat{s} \left(\hat{s} \left(\hat{s}+5\right)-6 \omega _0-1\right)+102
\omega _0+121}}{2 \left(\hat{s}-1\right) \sqrt{\frac{2 \omega _0}{3}+1}}$,%
\newline
$-\frac{3}{2 \left(\hat{s}-1\right)}+\frac{\sqrt{-8 \hat{s} \left(\hat{s}
\left(\hat{s}+5\right)-6 \omega _0-1\right)+102 \omega _0+121}}{2 \left(\hat{%
s}-1\right) \sqrt{\frac{2 \omega _0}{3}+1}},-\frac{6 F^{\prime }\left(\hat{s}%
\right)}{\hat{s}-1}\Big\}$.\newline
It is a saddle.

\item[$P_4(\hat{s})$:] $(x,y,z,\lambda,s)=\left(\frac{\sqrt{6}}{\hat{s}-1},%
\frac{\sqrt{\hat{s}+2}}{1-\hat{s}}, \frac{\sqrt{\hat{s} \left(\hat{s}%
+3\right)-6 \omega _0-7}}{1-\hat{s}},0 ,\hat{s}\right)$. Exists for

\begin{enumerate}
\item $\hat{s}=-2, \omega _0<-\frac{3}{2}$, or

\item $-2<\hat{s}<-1, \omega _0\leq \frac{1}{6} \left(\hat{s}^2+3 \hat{s}%
-7\right)$, or

\item $-1<\hat{s}<1,  -\frac{3}{2}<\omega _0\leq \frac{1}{6} \left(\hat{s}%
^2+3 \hat{s}-7\right)$, or

\item $-1\leq \hat{s}<1, \omega _0<-\frac{3}{2}$.
\end{enumerate}

The eigenvalues are \newline
$\Big\{\frac{3}{\hat{s}-1},6,-\frac{3}{2 \left(\hat{s}-1\right)}-\frac{\sqrt{%
-8 \hat{s} \left(\hat{s} \left(\hat{s}+5\right)-6 \omega _0-1\right)+102
\omega _0+121}}{2 \left(\hat{s}-1\right) \sqrt{\frac{2 \omega _0}{3}+1}}$,%
\newline
$-\frac{3}{2 \left(\hat{s}-1\right)}+\frac{\sqrt{-8 \hat{s} \left(\hat{s}
\left(\hat{s}+5\right)-6 \omega _0-1\right)+102 \omega _0+121}}{2 \left(\hat{%
s}-1\right) \sqrt{\frac{2 \omega _0}{3}+1}}, -\frac{6 F^{\prime }\left(\hat{s%
}\right)}{\hat{s}-1}\Big\}$.\newline
It is a saddle.

\item[$P_5(\hat{s})$:] $(x,y,z,\lambda,s)=\left(\frac{2 \sqrt{\frac{2}{3}}}{%
2 \omega _0+1},0,0,0,\hat{s}\right)$, where $F(\hat{s})=0$. \newline
Exists for

\begin{enumerate}
\item $\omega _0<-\frac{3}{2}$, or

\item $-\frac{5}{6}\leq \omega _0<-\frac{1}{2}$, or

\item $\omega _0>-\frac{1}{2}$.
\end{enumerate}

The eigenvalues are\\$\left\{\frac{2}{2 \omega _0+1},-\frac{2 \left(\hat{s}%
-1\right)}{2 \omega _0+1},-3,-\frac{6 \omega _0+5}{2 \omega _0+1},-\frac{4
F^{\prime }\left(\hat{s}\right)}{2 \omega _0+1}\right\}$. \newline
It is a sink for $\omega _0<-\frac{3}{2}, \hat{s}<1, F^{\prime }\left(\hat{s}%
\right)<0$. It is a saddle otherwise.

\item[$P_6(\hat{s})$:] $(x,y,z,\lambda,s)$\\$= \left(-\frac{\sqrt{\frac{2}{3}}
\left(\hat{s}+1\right)}{\hat{s}-2 \left(\omega _0+1\right)},\frac{\sqrt{%
\left(2 \omega _0+3\right) \left(-\hat{s} \left(\hat{s}+2\right)+6 \omega
_0+8\right)}}{\sqrt{3} \left(\hat{s}-2 \left(\omega _0+1\right)\right)},0,0,%
\hat{s}\right)$, where $F(\hat{s})=0$. Exists for

\begin{enumerate}
\item $\hat{s}<-1, \omega _0=\frac{1}{6} \left(\hat{s}-2\right) \left(\hat{s}%
+4\right)$, or

\item $\hat{s}<-1, \omega _0<\frac{1}{2}  \left(\hat{s}-2\right)$, or

\item $-1\leq \hat{s}\leq 2, \omega _0<-\frac{3}{2}$, or

\item $-1<\hat{s}<2, \frac{1}{6} \left(\hat{s}-2\right)  \left(\hat{s}%
+4\right)\leq \omega _0<\frac{1}{2} \left(\hat{s}-2\right)$, or

\item $\hat{s}>2, \omega _0=\frac{1}{6} \left(\hat{s}-2\right)  \left(\hat{s}%
+4\right)$, or

\item $\hat{s}>2, \omega _0<-\frac{3}{2}$.
\end{enumerate}

The eigenvalues are \newline
$\Big\{-\frac{\hat{s}+1}{\hat{s}-2 \omega _0-2},-\frac{2 \left(\hat{s}%
-1\right) \left(\hat{s}+1\right)}{\hat{s}-2 \omega _0-2},-\frac{\hat{s}^2+2
\hat{s}-6 \omega _0-8}{\hat{s}-2 \omega _0-2}, -\frac{\hat{s}^2+3 \hat{s}-6
\omega _0-7}{\hat{s}-2 \omega _0-2},\frac{2 \left(\hat{s}+1\right) F^{\prime
}\left(\hat{s}\right)}{\hat{s}-2 \omega _0-2}\Big\}$. \newline
It is a sink for $\omega _0<-\frac{3}{2}, \hat{s}>1, F^{\prime }\left(\hat{s}%
\right)<0$, or a saddle otherwise.

\item[$P_7(\hat{s})$:] $(x,y,z,\lambda,s)$\\
$=\left(-\frac{\sqrt{\frac{2}{3}}
\left(\hat{s}+1\right)}{\hat{s}-2 \left(\omega _0+1\right)},-\frac{\sqrt{%
\left(2 \omega _0+3\right) \left(-\hat{s} \left(\hat{s}+2\right)+6 \omega
_0+8\right)}}{\sqrt{3} \left(\hat{s}-2 \left(\omega _0+1\right)\right)},0,0,%
\hat{s} \right)$,\\ where $F(\hat{s})=0$. \newline
Exists for

\begin{enumerate}
\item $\hat{s}<-1, \frac{1}{2} \left(\hat{s}-2\right)<\omega _0<-\frac{3}{2}$%
, or

\item $\hat{s}<-1, \omega _0\geq \frac{1}{6} \left(\hat{s}-2\right)  \left(%
\hat{s}+4\right)$, or

\item $\hat{s}=-1, \omega _0>-\frac{3}{2}$, or

\item $-1<\hat{s}<2, \omega _0=\frac{1}{6} \left(\hat{s}-2\right)  \left(%
\hat{s}+4\right)$, or

\item $-1<\hat{s}<2, \omega _0>\frac{1}{2} \left(\hat{s}-2\right)$, or

\item $\hat{s}=2, \omega _0>0$, or

\item $\hat{s}>2, \omega _0\geq \frac{1}{6} \left(\hat{s}-2\right) \left(%
\hat{s}+4\right)$.
\end{enumerate}

The eigenvalues are\newline
$\Big\{-\frac{\hat{s}+1}{\hat{s}-2 \omega _0-2},-\frac{2 \left(\hat{s}%
-1\right) \left(\hat{s}+1\right)}{\hat{s}-2 \omega _0-2},-\frac{\hat{s}^2+2
\hat{s}-6 \omega _0-8}{\hat{s}-2 \omega _0-2}$,\\
$-\frac{\hat{s}^2+3 \hat{s}-6
\omega _0-7}{\hat{s}-2 \omega _0-2},\frac{2 \left(\hat{s}+1\right) F^{\prime
}\left(\hat{s}\right)}{\hat{s}-2 \omega _0-2}\Big\}$.\newline
It is always a saddle point.

\item[$P_8(\hat{s})$:] $(x,y,z,\lambda,s)=\left(\frac{\sqrt{3}-\sqrt{2
\omega _0+3}}{\sqrt{2} \omega _0},0,0,0,\hat{s}\right)$, where $F(\hat{s})=0$%
.\newline
Exists for $\omega _0>-\frac{3}{2}, \omega _0\neq 0$. \newline
The eigenvalues are \newline
$\Big\{-\frac{\sqrt{6 \omega _0+9}-3}{2 \omega _0},-\frac{\sqrt{6 \omega _0+9%
}-3}{2 \omega _0},\frac{6 \omega _0-\sqrt{6 \omega _0+9}+3}{\omega _0}$,%
\newline
$\frac{\hat{s} \left(\sqrt{6 \omega _0+9}-3\right)+6 \omega _0-2 \sqrt{6
\omega _0+9}+6}{2 \omega _0},\frac{\left(\sqrt{6 \omega _0+9}-3\right)
F^{\prime }\left(\hat{s}\right)}{\omega _0}\Big\}$.  \newline
It is a saddle.

\item[$P_9(\hat{s})$:] $(x,y,z,\lambda,s)=\left(\frac{\sqrt{3}+\sqrt{2
\omega _0+3}}{\sqrt{2} \omega _0},0,0,0,\hat{s}\right)$, where $F(\hat{s})=0$%
.\newline
Exists for $\omega _0>-\frac{3}{2}, \omega _0\neq 0$. \newline
The eigenvalues are \newline
$\Big\{\frac{\sqrt{6 \omega _0+9}+3}{2 \omega _0},\frac{\sqrt{6 \omega _0+9}%
+3}{2 \omega _0},\frac{6 \omega _0+\sqrt{6 \omega _0+9}+3}{\omega _0}$,%
\newline
$\frac{-\hat{s} \left(\sqrt{6 \omega _0+9}+3\right)+6 \omega _0+2 \sqrt{6
\omega _0+9}+6}{2 \omega _0},-\frac{\left(\sqrt{6 \omega _0+9}+3\right)
F^{\prime }\left(\hat{s}\right)}{\omega _0}\Big\}$.  \newline
It is a source for

\begin{enumerate}
\item $\hat{s}\leq 2, \omega _0>0, F^{\prime }\left(\hat{s}\right)<0$, or

\item $\hat{s}>2, \omega _0>\frac{1}{6} \left(\hat{s}-2\right) \left(\hat{s}%
+4\right),  F^{\prime }\left(\hat{s}\right)<0$.
\end{enumerate}

It is a sink for

\begin{enumerate}
\item $\hat{s}\leq 1, -\frac{5}{6}<\omega _0<0, F^{\prime }\left(\hat{s}%
\right)<0$, or

\item $1<\hat{s}<2, \frac{1}{6} \left(\hat{s}-2\right)  \left(\hat{s}%
+4\right)<\omega _0<0, F^{\prime }\left(\hat{s}\right)<0$.
\end{enumerate}
\end{enumerate}

As we see, the stability of these points depends on the character of the
zeros $\hat{s}$ of the function $F(s)$, and their first order derivative
evaluated at $\hat{s}$. The function $F(s)$ for the most common quintessence
potentials \cite{Escobar:2013js} is displayed in table \ref{fsform}.

The points $P_1$- $P_9$ were found in the previous three examples, for which
$s$ is a constant, such that $h$ is only a function of $\lambda$ that depends
on the choice of $W$, and $F$ is identically zero ( the problem can
be reduced in one dimension). When $F(s)$ is not trivial, the above
classification can be implemented straightforwardly, as for the specific
choices of $F$ in table \ref{fsform}.
The search of the equilibrium points with $\lambda\neq 0$ is not an easy
task, and the success on it depends crucially on the choice of $h(s,\lambda)$.
Indeed, for a given $h$, there are equilibrium points on the surface $x- 2
\lambda z (h(s,\lambda )-1)=0$. On this surface the existence conditions of
an equilibrium point are $\Big\{(x,y,z,\lambda, s)\in\mathbb{R}^5: y\geq 0, z\geq 0, \lambda \geq 0,2 \omega
_0+3\neq 0,  \notag \\
 z^2 \left(4 \lambda ^2 \omega _0 (h(s,\lambda )-1)^2+1\right)+y^2\leq 2
\sqrt{6} \lambda z (h(s,\lambda )-1)+1\Big\}.$

For example, the given $h(s,\lambda)\equiv 1$, we have the additional
equilibrium points $(x,y,z,\lambda,s)=\left(0,0,\sqrt{\frac{2}{3}}%
,2,s_c\right)$, where $h(s_c,2)=1$. For $h(s,\lambda)=1-\frac{1}{2\beta}$,
we have the additional points $P_{11}$- $P_{15}$ investigated in section \ref%
{sect:4.2}.

\begin{table*}
\caption{Cosmological parameters corresponding to the formulation in the
dilatonic frame as given by Eqs. \eqref{FE_jordan} for the equilibrium
points in the invariant set $\protect\lambda=0$ for arbitrary potentials.}
\label{Tab3b}
\begin{tabular*}{\textwidth}{@{\extracolsep{\fill}}lrrrl@{}}
\hline
Equil. & \multicolumn{1}{c}{$\Omega_2$} & \multicolumn{1}{c}{$w_1$} & \multicolumn{1}{c}{$w_2$} & \multicolumn{1}{c}{${w_{\text{tot}}}$}  \\
Points \\
\hline
$P_1$ & $\frac{s_c+1}{s_c-1}$ & $-1$ & $-1$ & $-1$  \\ \hline
$P_2(\hat{s})$ & $-\frac{2 \omega _0}{3}-1$ & $1$ & $1$ & $1$  \\ \hline
$P_3(\hat{s})$ & $\frac{\hat{s} \left(\hat{s}+3\right)-6 \omega _0-7}{\left(\hat{s}-1\right)^2}$ & $1$ & $1$ & $1$ \\ \hline
$P_4(\hat{s})$&  $\frac{\hat{s} \left(\hat{s}+3\right)-6 \omega _0-7}{\left(\hat{s}-1\right)^2}$ & $1$ & $1$ & $1$  \\ \hline
$P_5(\hat{s})$ & $\frac{\left(2 \omega _0+3\right) \left(6 \omega _0+5\right)}{3 \left(2 \omega _0+1\right){}^2}$ & $-1$ & $-1$ & $-1$ \\ \hline
$P_6(\hat{s})$ & $0$ & $-\frac{2 \left(\hat{s}^2-1\right)}{3 \left(\hat{s}-2 \left(\omega _0+1\right)\right)}-1$ & \text{Indeterminate} & $-\frac{2 \left(\hat{s}^2-1\right)}{3 \left(\hat{s}-2
   \left(\omega _0+1\right)\right)}-1$ \\ \hline
$P_7(\hat{s})$ & $0$ & $-\frac{2 \left(\hat{s}^2-1\right)}{3 \left(\hat{s}-2 \left(\omega _0+1\right)\right)}-1$ & \text{Indeterminate} & $-\frac{2 \left(\hat{s}^2-1\right)}{3 \left(\hat{s}-2
   \left(\omega _0+1\right)\right)}-1$ \\ \hline
$P_8(\hat{s})$ & $0$ & $\frac{3 \omega _0-\sqrt{6 \omega _0+9}+3}{3 \omega _0}$ & $-1$ & $\frac{3 \omega _0-\sqrt{6 \omega _0+9}+3}{3 \omega _0}$ \\ \hline
$P_9(\hat{s})$ & $0$ & $\frac{3 \omega _0+\sqrt{6 \omega _0+9}+3}{3 \omega _0}$ & $-1$ & $\frac{3 \omega _0+\sqrt{6 \omega _0+9}+3}{3 \omega _0}$ \\ 
\hline
\end{tabular*}
\end{table*}

\section{Discussion and Conclusions}
\label{Sect:7}
In this work the Brans-Dicke action has been considered in the cosmological
scenario of FLRW spacetime with spatially flat curvature; while a minimally coupled scalar
field was considered as a matter source. We show that this
action in the Einstein frame provides the dilatonic action integral and it
is equivalent to the $\sigma$-models. The method of the group invariant transformations, i.e., symmetries of
differential equations was applied in order to constraint the free functions of
the theory, and determine conservation laws for the gravitational field
equations. We found that for a family of potentials there exists a
Noetherian conservation law. From the admitted symmetries we derived the
zero-order invariants and we derived specific solutions for the field
equations which correspond to matter-like dominant eras. Additionally,
we have studied the stability of the equilibrium points of the dynamical system for specific and arbitrary potentials.

For the model 1,  corresponding to the formulation in the
dilatonic frame as given by Eqs. \eqref{FE_jordan} for $W(\protect\psi)=W_0
e^{k \protect\psi}$ and $V(\protect\varphi)=V_0 e^{l \protect\varphi}$, we have obtained the following  main results. The equilibrium for  $P_1$ corresponds to a solution with $w_{\text{tot}}=-1$.
We have proved that its center manifold is stable for $1<l\leq \frac{1}{16} \left(6 \omega _0+25\right)$. We show this solution is an attractor in the dilatonic frame but it is an intermediate accelerated solution $a \simeq e^{A t^p}, p:=\frac{2}{2+l}, \quad \frac{32}{57+6 \omega_0}<p<\frac{2}{3}, \;\text{as}\; t\rightarrow \infty$, and not a de Sitter solution. The exponent $p$ is reduced, in a particular case, to the exponent already found for the Jordan's and Einstein's frames by \cite{Cid:2015pja}. We have obtained some  equilibrium points, $P_2, P_3$ and $P_4$, that represent stiff solutions which  are saddle. The equilibrium point $P_5$, satisfies $w_{\text{tot}}=-1$. 
It is a sink for $l>-1, \omega _0<-\frac{3}{2}$ or a saddle otherwise.  The equilibrium point $P_6$,  corresponds to a solution where the energy density of the dilatonic field is dominant and the energy density of the quintessence field is negligible.  According to whether ${w_{\text{tot}}}:=\frac{2 \left(l^2-1\right)}{3 \left(l+2 \omega _0+2\right)}-1$ satisfies ${w_{\text{tot}}}<-1, {w_{\text{tot}}}=-1$ or $-1<{w_{\text{tot}}}<-\frac{1}{3}$ we have found it represents a phantom solution, a solution with $w_{\text{tot}}=-1$ or a quintessence solution. It is a sink for $l<-1, \omega _0<-\frac{3}{2}$. It is a saddle otherwise. Other equilibrium point as $P_8$ mimics a standard dark matter source with $0<w_{\text{tot}}<1$.
It is a saddle.  The equilibrium point $P_9$ corresponds to a solution where the energy density of the dilatonic field is dominant and the energy density of the quintessence field is negligible. Furthermore,  $w_1=\frac{3 \omega _0+\sqrt{6 \omega _0+9}+3}{3 \omega _0}$, $w_2=-1$, and $w_{\text{tot}}=\frac{3 \omega _0+\sqrt{6 \omega _0+9}+3}{3 \omega _0}$. That is, the second fluid behaves as a cosmological constant whereas the effective equation of state (of the total cosmic budget) is that of quintessence field for $-\frac{9}{8}<\omega_0<-\frac{5}{6}$,  a cosmological constant for $\omega_0=-\frac{5}{6}$ and the phantom field for $-\frac{5}{6}<\omega_0<0$.
It is a sink for $-2<l\leq -1, \frac{1}{6} (l-4) (l+2)<\omega _0<0$, or  $l>-1, -\frac{5}{6}<\omega _0<0$ (in both cases is a phantom attractor). It is a source for $l\leq -2, \omega _0>\frac{1}{6} (l-4) (l+2)$, or  $l>-2, \omega_0>0$ (and it behaves as a standard matter source then).
It is a saddle otherwise. Finally,
the equilibrium point $P_{10}$  is dominated by the quintessence field and the contribution of the dilatonic field to the total energy density is negligible. It satisfies $w_2=\frac{1}{3}$ and ${w_{\text{tot}}}=\frac{1}{3}$. This means that the corresponding cosmological solution mimics radiation. Interestingly, it is a saddle with a three dimensional stable manifold provided $\omega_0\geq \frac{45}{2}$.
These results illustrate the capabilities of the model.

For the second model, we have
$V\left( \varphi \right) =V_{0}e^{\left( \beta -1\right) \varphi }~\
,~W\left( \psi \right) =W_{0}\psi ^{2\beta }$. The particular parameters
where chosen to lead to Noether pointlike symmetries. For this model, we have the same results for the first nine equilibrium
points in Table \ref{Tab2b} , in section \ref%
{Sect:6.1}, and discussed before, by replacing $l=\beta-1, k=2\beta$. We have found the additional equilibrium
points $P_{11}-P_{15}$, whose stability and cosmological observables have to be evaluated
numerically.

We recall that
the points $P_1$- $P_9$ were found in the previous examples, under the assumption
$s$ is a constant such that $h$ is only a function of $\lambda$ that depends
on the choice of $W$, and $F$ is identically zero (that is, the problem can
be reduced in one dimension). When $F(s)$ is not trivial, the above
classification can be implemented straightforwardly, as for the specific
choices of $F$ in table \ref{fsform}.  The search of the equilibrium points with $\lambda\neq 0$ is not an easy
task, and the success on it depends crucially on the choice of $h(s,\lambda)$.
For example, given $h(s,\lambda)\equiv 1$, we have the additional
equilibrium points $(x,y,z,\lambda,s)=\left(0,0,\sqrt{\frac{2}{3}}%
,2,s_c\right)$, where $h(s_c,2)=1$. For $h(s,\lambda)=1-\frac{1}{2\beta}$,
we have the additional points $P_{11}$- $P_{15}$ investigated in section \ref%
{sect:4.2}.
A more complete study requires the specification of the free functions and this is far the scope of the present research.

A possible generalization on the context of scalar-tensor theories
will be interesting. In this respect, after dealing with two simple examples, we made the first steps to provide a complete dynamical system analysis
of dilatonic JBD cosmology keeping the potentials arbitrary, which is a major improvement since it
allows for  the extraction of information that is related to the foundations of the
cosmological model and not to the specific potentials forms. In particular, we apply an extended version of the
method of $f$-devisers \cite{Escobar:2013js,Fadragas:2013ina,delCampo:2013vka} - in the sense that it was developed for two free functions
such that additionally to the $f$-deviser we have an $h$-deviser. Using this approach, one first performs the  analysis without the need of a
priori  specification of the potentials forms. At the end, one just substitutes the
specific potential forms in the results, instead of having to repeat the whole dynamical
elaboration from the start. Therefore, the results are richer and more general,
revealing the full  capabilities of dilatonic JBD cosmology.

\section*{Acknowledgments}

This work was funded by Comisi\'on Nacional de Investigaci\'on Cient\'{\i}%
fica y Tecnol\'ogica (CONICYT) through FONDECYT Iniciaci\'on grant no.
11180126 (G. L., A. P.). A.P. thanks the University of Athens for the hospitality provided while part of this work was carried out. G. L. and L.V.A. thank to
Vicerrector\'ia de Investigaci\'on y Desarrollo Tecnol\'ogico at Universidad
Cat\'olica del Norte for financial support. G. L. thanks to M. Antonella Cid, and to Fabiola Arevalo for initial discussions about this subject.
 Ellen de los Milagros Fernández Flores is acknowledged for proofreading. 

\providecommand{\href}[2]{#2}\begingroup\raggedright

\endgroup

\end{document}